\documentclass[twocolumn]{aastex63}
\usepackage{amsmath}
\usepackage{mathtools}  
\usepackage{comment}
\usepackage{graphicx}
\usepackage{xcolor}
\usepackage{xparse,xcoffins}
\usepackage{appendix}

\shorttitle{MDN emulator for ISM emission}
\shortauthors{Yang et al.}

\ExplSyntaxOn
\NewCoffin\imagecoffin
\NewCoffin\labelcoffin

\keys_define:nn { miguel/label }
 {
  label   .tl_set:N = \l_miguel_label_tl,
  labelbox .bool_set:N = \l_miguel_label_box_bool,
  labelbox .default:n = true,
  fontsize .tl_set:N = \l_miguel_label_size_tl,
  fontsize .initial:n = \footnotesize,
  pos .choice:,
  pos/nw .code:n = \tl_set:Nn \l_miguel_label_pos_tl { left,up },
  pos/ne .code:n = \tl_set:Nn \l_miguel_label_pos_tl { right,up },
  pos/sw .code:n = \tl_set:Nn \l_miguel_label_pos_tl { left,down },
  pos/se .code:n = \tl_set:Nn \l_miguel_label_pos_tl { right,down },
  pos/n .code:n = \tl_set:Nn \l_miguel_label_pos_tl { hc,up },
  pos/w .code:n = \tl_set:Nn \l_miguel_label_pos_tl { left,vc },
  pos/s .code:n = \tl_set:Nn \l_miguel_label_pos_tl { hc,down },
  pos/e .code:n = \tl_set:Nn \l_miguel_label_pos_tl { right,vc },
  pos .initial:n = nw,
  unknown .code:n   = \clist_put_right:Nx \l_miguel_label_clist
                       { \l_keys_key_tl = \exp_not:n { #1 } }
 }
\clist_new:N \l_miguel_label_clist
\box_new:N \l_miguel_label_box
\box_new:N \l_miguel_label_image_box

\NewDocumentCommand{\xincludegraphics}{O{}m}
 {
  \group_begin:
  \tl_clear:N \l_miguel_label_tl
  \clist_clear:N \l_miguel_label_clist
  \keys_set:nn { miguel/label } { #1 }
  \tl_if_empty:NTF \l_miguel_label_tl
   {
    \miguel_includegraphics:Vn \l_miguel_label_clist { #2 }
   }
   {
    \SetHorizontalCoffin\imagecoffin
     {
      \miguel_includegraphics:Vn \l_miguel_label_clist { #2 }
     }
    \SetHorizontalCoffin\labelcoffin
     {
      \raisebox{\depth}
       {
        \bool_if:NTF \l_miguel_label_box_bool
         { \fcolorbox{white}{white}{\l_miguel_label_size_tl\l_miguel_label_tl} }
         { \l_miguel_label_size_tl\l_miguel_label_tl }
       }
     }
    \SetVerticalPole\imagecoffin{left}{3pt+\CoffinWidth\labelcoffin/2}
    \SetVerticalPole\imagecoffin{right}{\Width-3pt-\CoffinWidth\labelcoffin/2}
    \SetHorizontalPole\imagecoffin{up}{\Height-3pt-\CoffinHeight\labelcoffin/2}
    \SetHorizontalPole\imagecoffin{down}{3pt+\CoffinHeight\labelcoffin/2}
    \use:x{\JoinCoffins\imagecoffin[\l_miguel_label_pos_tl]\labelcoffin[vc,hc]} 
    \TypesetCoffin\imagecoffin
   }
   \group_end:
 }
\NewDocumentCommand{\setlabel}{m}
 {
  \keys_set:nn { miguel/label } { #1 }
 }

\cs_new_protected:Nn \miguel_includegraphics:nn
 {
  \includegraphics[#1]{#2}
 }
\cs_generate_variant:Nn \miguel_includegraphics:nn { V }

\ExplSyntaxOff

\begin{document}

\title{A New Framework for ISM Emission Line Models: Connecting Multi-Scale Simulations Across Cosmological Volumes}

\correspondingauthor{Shengqi Yang}
\email{syang@carnegiescience.edu}

\author{Shengqi Yang}
\affiliation{Carnegie Observatories, 813 Santa Barbara Street, Pasadena, CA 91101, U.S.A}
\affiliation{Los Alamos National Laboratory, Los Alamos, NM 87545, USA}

\author{Adam Lidz}
\affiliation{Department of Physics and Astronomy, University of Pennsylvania, 209 South 33rd Street, Philadelphia, PA 19104, USA}

\author{Andrew Benson}
\affiliation{Carnegie Observatories, 813 Santa Barbara Street, Pasadena, CA 91101, U.S.A}

\author{Yizhou Zhao}
\affiliation{UCLA Center for Vision, Cognition, Learning, and Autonomy}

\author{Hui Li}
\affiliation{Los Alamos National Laboratory, Los Alamos, NM 87545, USA}

\author{Amelia Zhao}
\affiliation{Pomona College, Claremont, CA 91711, USA}

\author{Aaron Smith}
\affiliation{Department of Physics, The University of Texas at Dallas, Richardson, Texas 75080, USA}

\author{Yucheng Zhang}
\affiliation{Peng Cheng Laboratory, Shenzhen, Guangdong 518066, China}

\author{Rachel Somerville}
\affiliation{Center for Computational Astrophysics, Flatiron Institute, New York, NY 10010, USA}

\author{Anthony Pullen}
\affiliation{Center for Cosmology and Particle Physics, Department of Physics, New York University, 726 Broadway, New York, NY 10003, USA}

\author{Hui Li}
\affiliation{Department of Astronomy, Tsinghua University, Beijing 100084, China}



\begin{abstract}
The JWST and ALMA have detected emission lines from the ionized interstellar medium (ISM) in some of the first galaxies at $z \gtrsim 6$.  These measurements present an opportunity to better understand galaxy assembly histories and may allow important tests of state-of-the-art galaxy formation simulations. It is challenging, however, to model these lines in their proper cosmological context. In order to meet this challenge, we introduce a novel sub-grid line emission modeling framework. The framework uses the high-$z$ zoom-in simulation suite from the Feedback in Realistic Environments (FIRE) collaboration. The line emission signals from HII regions within each simulated FIRE galaxy are modeled using the semi-analytic \textsc{HIILines} code. A machine learning approach is then used to determine the conditional probability distribution for the line luminosity to stellar-mass ratio from the HII regions around each simulated stellar particle. This conditional probability distribution can then be applied to predict the line luminosities around stellar particles in lower resolution, yet larger volume cosmological simulations. As an example, we apply this approach to the IllustrisTNG simulations at $z=6$. The resulting predictions for the [OII], [OIII], and Balmer line luminosities as a function of star-formation rate (SFR) agree well with current observations. Our predictions differ, however, from related work in the literature which lack detailed sub-grid ISM models. This highlights the importance of our multi-scale simulation modeling framework. Finally, we provide forecasts for future line luminosity function measurements from the JWST and quantify the cosmic variance in such surveys. 
\end{abstract}

\section{Introduction}
The Atacama Large Millimeter/Submillimeter Array (ALMA) and the James Webb Space Telescopes (JWST) have detected multiple emission lines from some of the first galaxies, which formed less than a billion years after the Big Bang (e.g. \cite{2016Sci...352.1559I,2017ApJ...837L..21L,2018Natur.557..392H,2019MNRAS.487L..81L,2019PASJ...71...71H,2020ApJ...896...93H,2022MNRAS.515.1751W,2022arXiv221202890H,2023MNRAS.518..425C,2023arXiv230308149S,2023arXiv230603120L}). Excitingly, the pace of discovery here is expected to accelerate over the next few years. In addition, SPHEREx will soon be launched and it will, among other measurements, carry out line-intensity mapping (LIM) surveys. This will uniquely complement the bright galaxy detections from the JWST and ALMA \citep{2014arXiv1412.4872D}. Among the brightest nebular lines currently detected are collisionally-excited [OII] and [OIII] lines, and hydrogen Balmer series recombination lines. These lines are sensitive to the intensity of the local ionizing radiation and the properties of the surrounding interstellar gas, including its metallicity, density, and temperature. 
The ALMA, JWST, and LIM measurements can then provide new insights into galaxy formation at high redshift, and can, in-principle, provide strong tests of state-of-the-art galaxy formation models as implemented in numerical simulations. \par

In order to best interpret the current and upcoming data, reliable emission line models across cosmological volumes are required. For example, one important goal is to use the observations to quantify correlations between the gas-phase metallicities and stellar masses of galaxies (the mass-metallicity relation, or ``MZR''). The amplitude, shape, scatter, and redshift evolution of this correlation are partly shaped by poorly understood feedback processes and the MZR hence provides an important empirical anchor for galaxy formation models \citep{2004ApJ...613..898T,2006ApJ...647..970L,2008A&A...488..463M,2009MNRAS.398.1915M,2013ApJ...771L..19Z,2021ApJ...919..143H}. However, to robustly extract this from current data, one requires models to infer gas-phase metallicities from the observable [OIII], [OII], and Balmer line luminosities, (and unbiased stellar mass estimates, although this is not our focus here).  Ideally, this necessitates modeling variations in the properties of the emitting HII regions across each galaxy, as well as accounting for galaxy-to-galaxy fluctuations in the ISM properties. Furthermore, the model must span the observed range in stellar mass. Hence, to best extract the MZR from current data, among other goals, one needs to capture HII-region scale physics across cosmological volumes.\par 

Although multiple strategies have been proposed for post-processing ISM emission lines on top of cosmological simulations, they are not ideal for making detailed comparisons between simulations and observations. The problem is two-fold. First, large volume cosmological simulations currently lack the resolution to capture the multi-phase ISM. Instead, a simplified two-phase description is used to characterize the sub-grid ISM, with `hot' and 'cold' phases evolving according to effective equations of state \citep{Springel2003}. This approach is inadequate for modeling the line emission from HII regions, which is partly shaped by the radiation field, metallicity, density, and temperature on sub-grid length scales. Second, it is expensive to perform line emission calculations across many millions of emitting regions using the \textsc{Cloudy} code \citep{2017RMxAA..53..385F}, especially if one properly accounts for variations in the stellar radiation field and in the local ISM properties. As a compromise, approximations are often adopted such as assuming a constant
stellar radiation field and/or gas density across all emitting regions (e.g. \cite{2017MNRAS.472.2468H,2018MNRAS.481L..84M,2021MNRAS.504.4472C,2022MNRAS.514.3857K,2023MNRAS.526.3610H,2023ApJ...953..140N}). Although such simplifications help make the problem tractable, it is unclear how accurate they are. The simple approach also discards information from the simulations regarding the simulated star-formation histories, which should ideally be incorporated into the line emission modeling and lead to region-to-region variations in the ionizing spectral amplitude and shape. On the other hand, there are expensive small-volume cosmological simulations that do, at least partly, resolve the multi-phase ISM. These calculations, however, generally simulate volumes too small to capture rare, bright galaxies, yet these sources are often the most accessible with current observations. As several representative examples from current state-of-the-art work in the literature, we note that \cite{2022MNRAS.514.3857K}
assumes a fixed HII region gas density across all galaxies and only accounts for stellar particles younger than 10 Myr. Likewise, \cite{2023MNRAS.526.3610H} adopt a constant gas density and assume all galaxies in the TNG simulation suite have identical stellar spectra. Even the more ambitious approach developed in \cite{2022MNRAS.510.5603K} only models line emission signals for a sub-set of simulation cells using \textsc{Cloudy} and then relies on a machine learning technique to extrapolate predictions across all cells. 
Moreover, the SPHINX simulations employed in \cite{2022MNRAS.510.5603K} span relatively small volumes and hence lack galaxies with star formation rates (SFRs) that are high enough to compare directly with many current ALMA 
measurements.\par
\begin{figure*}
    \centering
    \includegraphics[width=0.95\textwidth]{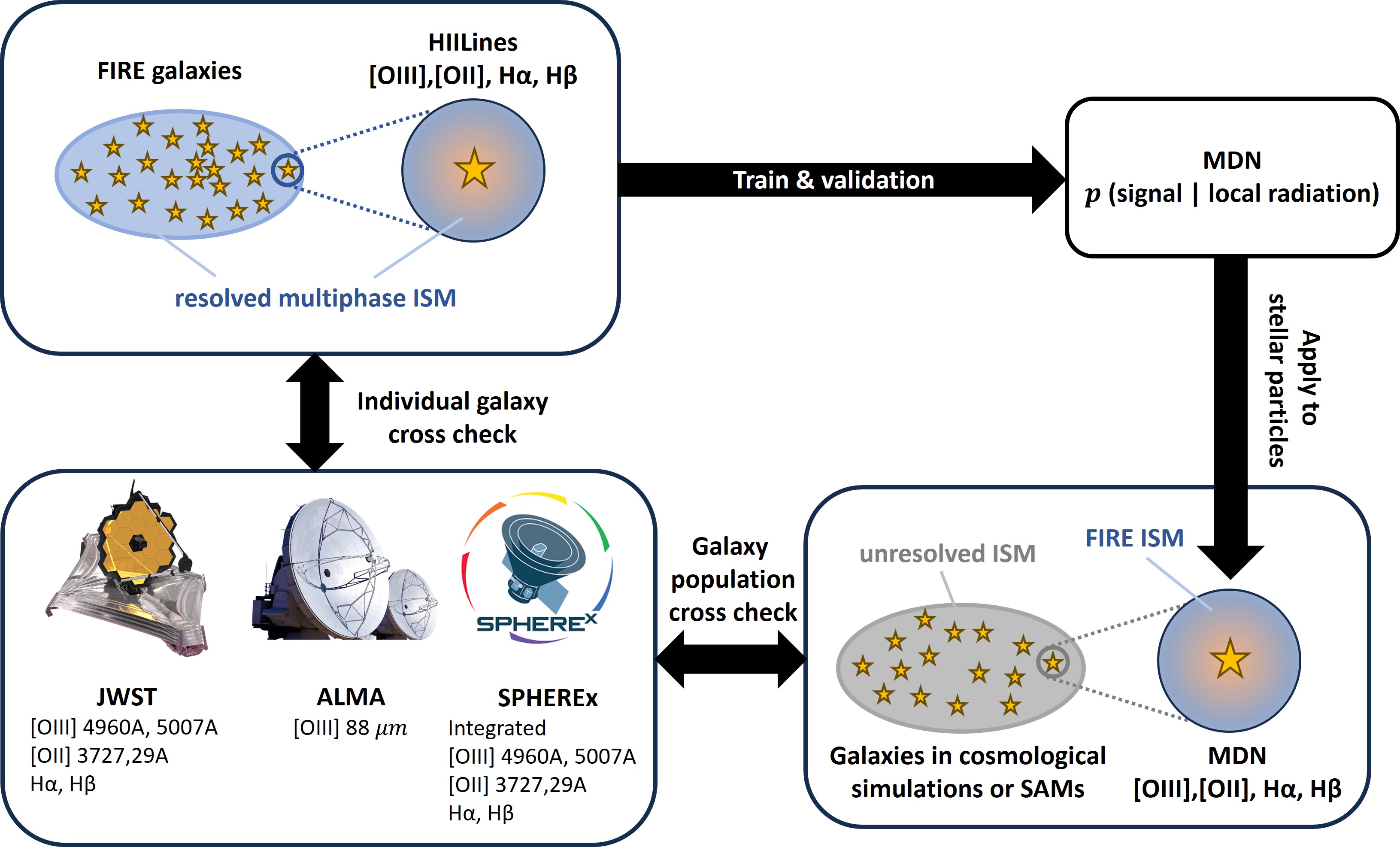}
    \caption{A schematic showing the main goal of this work. First, we use the \textsc{HIILines} code to model HII region emission line signals on top of FIRE zoom-in simulations of high redshift galaxies (top left). Next, an MDN approach is applied to the FIRE models to determine the joint cPDFs for the luminosities across several lines of interest, conditioned on the coarse-grained stellar particle properties (top right). This is then used to populate line emission signals around the stellar particles in lower resolution, yet larger volume, simulations (bottom right). That is, the cPDFs effectively provide a FIRE-informed sub-grid model for the fine-grained HII region and line emission properties, given the surrounding coarse-grained stellar particle properties. 
    The resulting models can then be compared in detail with current and upcoming [OIII], [OII], and hydrogen Balmer line observations from the JWST, ALMA, and SPHEREx ({\em bottom left}).}\label{fig:framework}
\end{figure*}

In this work we propose a new framework for post-processing ionized ISM emission lines on top of cosmological simulations which may improve upon some of the above limitations. This framework combines the high-$z$ FIRE zoom-in galaxy simulations \citep{2018MNRAS.478.1694M,2019MNRAS.487.1844M,2020MNRAS.493.4315M}, the mixture density network (MDN) architecture \citep{Bishop1994MixtureDN}, and an efficient ionized ISM emission line simulation method---\textsc{HIILines} \citep{2023arXiv231209213Y}. The main scientific goal of this novel hybrid framework is summarized in Figure~\ref{fig:framework}. We aim to use the MDN machine learning architecture to extract information about the small-scale structure of the ISM and the resulting emission line signals, conditioned on the properties of the local ionizing sources, using simulated FIRE high-$z$ galaxies that (partly) resolve individual HII regions. We then apply the the conditional probability distribution functions (cPDFs) to the stellar particles in cosmological simulations to predict the unresolved line emission around them. 
In this manner we can robustly post-process HII region emission line signals among galaxies in cosmological simulations, assuming a small-scale ISM environment identical to that in the FIRE zoom-in galaxies. That is, our model combines the population-level statistics of a cosmological simulation of galaxy formation with the small-scale ISM physics captured in FIRE. It is flexible enough for us to explore how varying some of the assumptions here impacts our final results. 
Ultimately, our framework will help fully exploit revolutionary new data sets from ALMA, JWST, and LIM surveys, across multiple wavebands, extracting their implications for early galaxy formation in its full cosmological context.\par
\par
The trained model is publicly available at \url{https://github.com/Sheng-Qi-Yang/GMDN}, and is applicable to any numerical or semi-analytical galaxy formation simulation with spatial resolution no finer than that of FIRE. The plan of this paper is as follows. In Section~\ref{sec:HIILines} we introduce the FIRE zoom-in galaxies and \textsc{HIILines}, which are used to generate the training and validation datasets for this work. Section~\ref{sec:MDN} introduces the MDN architecture and the model training method. Section~\ref{sec:MDNverify} explains how the trained MDN can be applied to stellar particles in simulated galaxies. We verify our MDN model by showing that it successfully reproduces the 1D distributions, 2D correlations, and galaxy-wide total line emission signals for the 22 FIRE galaxies. We also test the assumption that the line luminosity given stellar particle property cPDFs from these 22 FIRE galaxies are representative of more general galaxy populations. In Section~\ref{sec:TNG} we apply the trained MDN to well-resolved galaxies in the TNG50, TNG100, and TNG300 simulation suites at $z=6$. We then compare our models with current observations, considering multiple different statistical measures, and contrast our results with other modeling approaches in the literature. We summarize the strengths, limitations, and future prospects of our emission line modeling framework in Section~\ref{sec:summary}.

\section{Tools and Methods}\label{sec:HIILines}
\subsection{FIRE high-$z$ galaxies post-processed by \textsc{HIILines}}
The 22 central galaxies at $z=6$ from the FIRE project, made publicly available by \cite{2018MNRAS.478.1694M,2019MNRAS.487.1844M,2020MNRAS.493.4315M}, cover a wide range of stellar mass $M_{*,gal}\sim10^6M_\odot-10^{10}M_\odot$, gas-phase metallicity $Z\sim10^{-3}Z_\odot-0.6Z_\odot$, and morphology. Since the FIRE zoom-in simulations can resolve the multi-phase ISM at $\sim10$ parsec spatial resolution, which is comparable to the size of individual HII regions, the ISM environment simulated by FIRE is more reliable than those predicted by large-volume cosmological simulations. The FIRE zoom-in galaxy simulations can hence be combined with spectral synthesis calculations to model ISM emission line signals.\par
\begin{figure}
    \centering
    \includegraphics[width=0.49\textwidth]{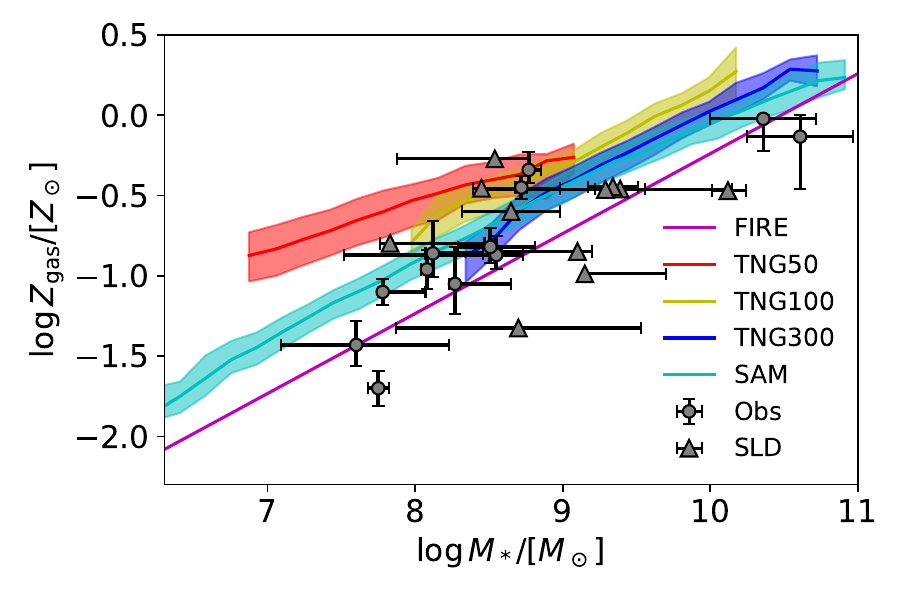}
    \caption{MZR at $z=6$ from FIRE, TNG, and the Santa Cruz SAM, as compared to direct $T_\mathrm{e}$ measurements. The metallicity of each simulated galaxy is defined as the SFR-weighted gas-phase metallicity. The colored bands give the 1$\sigma$ dispersion around the MZRs for the TNG and SAM models.  
    The FIRE high-$z$ suite contains only 22 central galaxies, and we therefore only show the best fit MZR in this case. Observational measurements are shown by the grey points and error bars. The TNG MZR lies above the observations at $M_*\lesssim10^{8.5} M_\odot$.}\label{fig:MZR}
\end{figure}

As an example, Figure~\ref{fig:MZR} compares the MZR at $z=6$ among the FIRE high-$z$ suite \citep{2018MNRAS.478.1694M,2019MNRAS.487.1844M,2020MNRAS.493.4315M}, TNG50, TNG100, TNG300 \citep{2018MNRAS.473.4077P,2018MNRAS.480.5113M,2018MNRAS.475..624N,2018MNRAS.475..676S,2018MNRAS.477.1206N,2018MNRAS.475..648P,2019MNRAS.490.3234N,2019MNRAS.490.3196P,2019ComAC...6....2N}, and the Santa Cruz semi-analytic galaxy formation model (SAM) \citep{1999MNRAS.310.1087S,2008MNRAS.391..481S,2012MNRAS.423.1992S,2014MNRAS.444..942P,2014MNRAS.442.2398P,2015MNRAS.453.4337S}, together with direct metallicity measurements from JWST \citep{2023ApJS..269...33N,2023MNRAS.518..425C} and ALMA \citep{2020ApJ...896...93H,2020MNRAS.499.3417Y} at $z\geq6$. When the [OIII] auroral line is undetected, metallicities are estimated through the O32+R23 strong line diagnostics using the method proposed by \cite{2023arXiv231209213Y} (triangular points) \citep{2023ApJS..269...33N}. In each case, the metallicity of a simulated galaxy is defined as the SFR-weighted gas-phase metallicity of oxygen averaged over all gas particles, which is close to the metallicity constrained by the ISM emission line measurements \citep{2023MNRAS.525.5989Y}. Although the scatter among the observational data points is large, the gas-phase metallicities of the TNG galaxies lie above the data for $M_*\lesssim10^{8.5}M_\odot$, while the FIRE and SAM MZRs appear in better agreement with current measurements. Since the small-scale ISM properties in FIRE, or at least the gas-phase metallicities, are in reasonable agreement with observations across a broad range of stellar masses, we use FIRE to anchor our modeling framework on small scales. 
\par 
The next tool in our modeling is the \textsc{HIILines} code from 
\cite{2023MNRAS.525.5989Y} which yields efficient [OIII], [OII], H$\alpha$, and H$\beta$ line emission predictions, post-processed on top of zoom-in galaxy formation simulations. \textsc{HIILines} assumes that every stellar particle in a simulated galaxy sources an independent HII region. The stellar radiation field is modeled using the Flexible Stellar Population Synthesis code (FSPS; \cite{2009ApJ...699..486C,2010ApJ...712..833C}), which takes the stellar particle age, metallicity, and mass as inputs. The gas density and metallicity of each HII region are assumed to be uniform and are estimated through averaging over the 32 nearest gas particle neighbors around the stellar particle of interest. Approximating the gas density and metallicity as uniform across each HII region is necessary because of the limited resolution of the FIRE simulations. This may lead to a mild underestimation of the rest-frame infrared [OIII] line luminosities owing to collisional de-excitations in unresolved density peaks \citep{2020MNRAS.499.3417Y}. However, the uniform approximation should be better for the rest-frame [OIII] optical lines and hydrogen Balmer lines due to their higher critical densities. An empirically-motivated model is used to describe the gas temperature since this quantity is not reliably modeled in the FIRE simulations. This model exploits the fact that the temperature of the HII region gas is well-correlated with its gas-phase metallicity. 

\par

Specifically, we assume that the temperature of all of the OIII and OII gas in a given galaxy is constant -- although the OIII and OII regions generally differ from each other in temperature --  and that these temperatures are determined by the
$Q_\mathrm{HI}$-weighted gas-phase metallicity of the galaxy, $Z_Q$. Here $Q_\mathrm{HI}$ is the ionizing photon generation rate of the stellar particles. The OIII temperature versus metallicity relation is fit to high-$z$ direct $T_\mathrm{e}$ measurements, while the OII temperature is fit to local measurements \citep{2020A&A...634A.107Y}. \textsc{HIILines} then applies a semi-analytic line emission model, first proposed by \cite{2020MNRAS.499.3417Y}, to compute the [OIII], [OII], H$\alpha$, and H$\beta$ line luminosities of each HII region. The semi-analytic model is $\sim\times1000$ times faster than numerical spectral synthesis codes such as \textsc{Cloudy}, and is therefore suitable for post-processing millions of line emitters with different ionizing radiation fields and gas properties. Thanks to its high computational efficiency, \textsc{HIILines} is currently the only model that accounts for spatial variations
in the local gas density, metallicity, and incident radiation spectrum, along with their scatter across
the HII regions within simulated galaxies. The simulated [OIII], [OII], H$\beta$ line luminosities and line ratios among the 22 FIRE high-$z$ galaxies are in excellent agreement with recent ALMA and JWST measurements at $z\gtrsim6$.\par
In this work we suppose that the ISM density and metallicity distributions within these 22 FIRE galaxies are representative and apply to any galaxy with similar mass/metallicity, regardless of their morphology. In other words, we assume that the ISM line luminosity cPDFs extracted from these FIRE zoom-in galaxies are universally applicable, and can be applied to arbitrary members of the simulated galaxy populations across large-volume cosmological simulations. This is a strong assumption, yet it lies at the foundation of our method for constructing sub-grid ISM line emission models. We will quantitatively test this assumption in Section~\ref{subsec:assumptionTest}.\par

\subsection{Datasets}
Each stellar particle in FIRE has a known radiation field, determined through stellar population synthesis calculations, and is assumed to source an independent HII region. 
We then model the luminosities for 13 lines emitted by each HII region, covering [OIII] 88 $\mu$m, 52 $\mu$m, 4960, 5007, 4364 \AA, [OII] 3727, 3729, 2471.0, 2471.1, 7322, 7332\AA, H$\alpha$, and H$\beta$ lines. Our goal is to determine the following joint cPDF:
\begin{equation}\label{eq:cPDF}
    \phi=p(\log(L^\mathrm{[OIII],[OII],\mathrm{H}\alpha,\mathrm{H}\beta}/m_*)|\log t,\log Z_*,\log M_\mathrm{*})\,.
\end{equation}
Here $L^\mathrm{[OIII],[OII],\mathrm{H}\alpha,\mathrm{H}\beta}$ describes the luminosities of the 13 [OIII], [OII], and hydrogen recombination lines emitted by an HII region in a FIRE galaxy, while $m_*$ denotes the mass of the stellar particle that sources the particular HII region. We do not fit for the line luminosity distribution directly, but instead fit for the ratio of line luminosity to stellar mass because we want the trained model to be applicable to simulations of arbitrary resolution.  The quantities $t$ and $Z_*$ are the age and metallicity of the stellar particle, respectively. Given a stellar population synthesis model, these two input parameters fully determine the shape of the local ionizing radiation field. Following \cite{2023MNRAS.525.5989Y}, we model stellar radiation using the FSPS code.  Finally, $M_*$ is the total stellar mass of a FIRE galaxy. We introduce $M_*$ as a third input parameter because it provides information about the gas temperature. As introduced in Section~\ref{sec:HIILines}, \textsc{HIILines} estimates the galaxy-wide HII region temperature based on $Z_Q$. Since $Z_Q$ is somewhat expensive to calculate directly, we instead use the MZR as a short-cut here: given the correlation between $Z_Q$ and $M_*$ in FIRE -- which agrees well with observations (Figure~\ref{fig:MZR}) -- $M_*$ implies information about $Z_Q$, which in turn specifies the model gas temperatures. \par 
The 22 FIRE galaxies contain $3.3\times10^6$ HII regions in total, sourced by stellar particles younger than 100 Myr. We randomly sample 80\% of the HII regions to form the training dataset. The remaining 20\% of the HII regions will be used for model validation. We do not account for HII regions sourced by extremely old stellar populations in this work as they are faint and contribute negligibly to the total galaxy-wide line luminosities (e.g. \cite{2023MNRAS.525.5989Y}).\par 

\begin{figure*}
    \centering
    \includegraphics[width=0.95\textwidth]{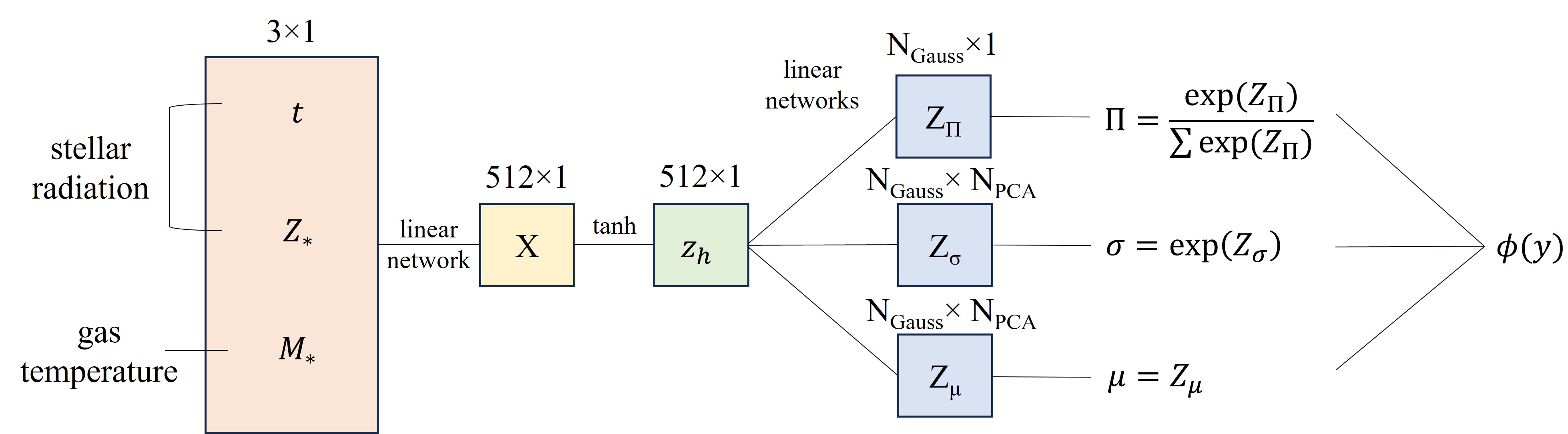}
    \caption{GMDN work flow. Given the three input parameters for each stellar particle: age, $t$; metallicity, $Z_*$; and galaxy total stellar mass, $M_*$, GMDN constructs a Gaussian Mixture Model (GMM) with
    weight, $\Pi$, standard deviation, $\sigma$, and mean, $\mu$, to describe the output, $y$. Here $y$ is the standardized line luminosity versus stellar particle mass ratio $\log(L/m_*)$ after PCA transformation. GMDN then randomly samples from the GMM to generate line signals for any given stellar particle. The advantage of this framework is that the local ISM environment from FIRE is encoded in the free parameters of the linear networks during the model training. The GMDN can then generate line luminosity models
    given only our three input parameters (stellar age, metallicity, and the total stellar mass of a galaxy), without requiring further information regarding the surrounding ISM. 
    }\label{fig:MDN}
\end{figure*}\par

\section{Galaxy-MDN (GMDN)}\label{sec:MDN}
In this section we combine the FIRE data post-processed by \textsc{HIILines} with a machine learning architecture named MDN to extract the connection between stellar particle properties and the corresponding HII region line signals. The combined pipeline here will be referred as Galaxy-MDN (GMDN).\par 
MDN is a machine learning architecture that fits the cPDF of the training data with a Gaussian mixture model (GMM, i.e., by a sum of multiple multi-variate Gaussian distributions), conditioned on the input variables \citep{Bishop1994MixtureDN}. Through randomly sampling over the GMM, MDN serves as a generative model and provides outputs with cPDFs approximately matching the distribution of the training data.\par
We summarize the overall flow of the GMDN in Figure~\ref{fig:MDN}. As introduced in Eq~\ref{eq:cPDF}, each stellar particle is assumed to be surrounded by an HII region and carries a 3-dimensional input array: $x=\{\log t, \log Z_*, \log M_*\}$. This input array, after standardization, will first go through a network layer for linear transformation:
\begin{equation}
    X = W_\mathrm{in}x+b_\mathrm{in}\,.
\end{equation} 
Here $W_\mathrm{in}$ and $b_\mathrm{in}$ are matrices of size $512\times3$ and $512\times1$ respectively. The size of the network is a hyper-parameter. We select 512 due to its balance between computational efficiency and memory usage, which is well-suited for our GPU computing framework. We then add non-linearity to the network result:
\begin{equation}
    z_h=\tanh(X)\,.
\end{equation}
Introducing non-linearity between each layer of linear transformation is necessary. Without this step, the combination of multiple layers of linear transformation is effectively only a single layer transformation, and the free parameters introduced by each network layer become redundant.\par 
Several linear networks are then used to compute the weight, mean, and standard deviation for each Gaussian distribution that makes up the GMM. We will use $N_\mathrm{Gauss}$ separate Gaussian distributions to describe the cPDF of the training data set. Ideally, each Gaussian should be 13-dimensional, corresponding to the 13 [OIII], [OII], and hydrogen recombination lines covered by \textsc{HIILines}. However, the luminosity distributions of these 13 lines are not independent of each other. Instead, they are highly positively correlated. In this case, the GMM requires a covariance matrix to capture the correlations among different lines. In order to simplify the GMM calculations, we perform a principle component analysis (PCA): we determine the eigenvectors of the covariance matrix and rank them according to their eigenvalues. 
We find that the first 4 PCA components account for more than 99\% of the total data variance and that the covariance matrix
can therefore be well approximated by including only the first $N_\mathrm{PCA}=4$ eigenmodes. 
In the eigenmode basis, in which the covariance matrix is diagonal, the GMM may be described as:
\begin{equation}\label{eq:GMM}
\begin{split}
    \phi(y)=&\sum_{k=1}^{N_\mathrm{Gauss}}\dfrac{\Pi_k}{(2\pi)^{N_\mathrm{PCA}/2}\prod_{i=1}^{N_\mathrm{PCA}}\sigma_k^i(x)} \\
    &\times\exp\left(-\dfrac{1}{2}\sum_{i=1}^{N_\mathrm{PCA}}\dfrac{(y^i-\mu_k^i(x))^2}{(\sigma_k^i(x))^2}\right)\,.
\end{split}
\end{equation}
Here $y^i$ describes the $i$th PCA eigenmode: these are suitable linear combinations of the line luminosities, $\log(L/m_*)$. 
\par

The weight of each Gaussian component $\Pi$ is computed through:
\begin{equation}
    Z_\Pi=W_\Pi z_h+b_\Pi\,, \Pi=\dfrac{\exp(Z_\Pi)}{\Sigma\exp(Z_\Pi)}\,.
\end{equation}
Here $W_\Pi$ and $b_\Pi$ are matrices of size $N_\mathrm{Gauss}\times512$ and $N_\mathrm{Gauss}\times1$, respectively.
The mean and standard deviation of each Gaussian for each PCA component $\mu$ and $\sigma$ are computed through:
\begin{equation}
    Z_\mu=W_\mu z_h+b_\mu\,, \mu=Z_\mu\,,
\end{equation}
and
\begin{equation}
    Z_\sigma=W_\sigma z_h+b_\sigma\,, \sigma=\exp(Z_\sigma)\,.
\end{equation}
Here $W_\mu$ and $W_\sigma$ are matrices of size $N_\mathrm{Gauss}\times512$. The quantities $b_\mu$ and $b_\sigma$ are matrices of size $N_\mathrm{Gauss}\times1$. There are, in total, 8 of these linear networks since each Gaussian is 4-dimensional.
The full GMDN contains in total 
\begin{equation}\label{eq:dimtheta}
    \mathrm{dim}\left(\theta\right) = 2048+4617 N_\mathrm{Gauss}
\end{equation}
free parameters. The number of Gaussian distributions used to build the GMM, $N_\mathrm{Gauss}$, is another hyper-parameter.\par
\subsection{Loss function}
Our goal is to train the GMDN so that, given the age and metallicity of a stellar particle from any simulation, together with the galaxy-wide stellar mass, the GMDN can generate random line luminosities with distributions and correlations identical to the FIRE simulation results. First, the PCA transfers all standardized FIRE HII region line luminosity-to-stellar mass ratios from 13D to 4D. We then define the first part of the loss function as the following:
\begin{equation}
\mathcal{L}_1=-\langle\omega(t_j)\ln\phi(y_\mathrm{FIRE}^j)\rangle_{j=1}^{N_\mathrm{HII}}\,,
\end{equation}
for both the training and validation datasets. Here $\langle\rangle_{j=1}^{N_\mathrm{HII}}$ denotes the mean among all HII regions in the training or validation dataset, while $\phi(y)$ is defined in Eq~\ref{eq:GMM}. The quantity $\omega(t_j)$ is a normalized weight for an HII region, which is determined by the stellar particle age. Specifically, we define the weight of an HII region to be inversely proportional to the stellar particle age PDF. We make this choice because young stellar populations are rare, yet they source the brightest HII regions, which are major contributors to the galaxy-wide total line luminosities. On the other hand, old stellar populations dominate the galaxy-wide stellar mass, yet they source faint HII regions and make negligible contributions to the total line luminosities. By introducing $\omega(t)$ in the loss function, we assign higher weights to the young stellar particles to better capture the total, galaxy-wide luminosities.\par
To ensure that the strong line diagnostics generated by our GMDN are close to the FIRE simulation results, we draw random samples from $\phi$ and do an inverse PCA transfer to get 13D outputs.  We then define the second part of the loss function as:
\begin{equation}
    \mathcal{L}_2=\sum_{l=1}^{5}\left\langle\omega(t_j)\times|\ln R^l_\mathrm{MDN}-\ln R^l_\mathrm{FIRE}|\right\rangle_{j=1}^{N_\mathrm{HII}}\,,
\end{equation}
where $R^1$ to $R^5$ denote the following strong line diagnostics: $L^\mathrm{[OIII]}_{5007}/L^\mathrm{[OIII]}_{4364}$, $L^\mathrm{[OII]}_{3729}/L^\mathrm{[OII]}_{3727}$, $L^\mathrm{[OIII]}_{88\mu m}/L^\mathrm{[OIII]}_{52\mu m}$, $L^\mathrm{[OIII]}_{5007}/L_\mathrm{H\beta}$, $L_\mathrm{H\alpha}/L_\mathrm{H\beta}$. The total loss function is:
\begin{equation}\label{eq:loss}
    \mathcal{L}=\mathcal{L}_1+\mathcal{L}_2\,.
\end{equation}
We introduce line ratio constraints into the loss function because these five line ratios are sensitive diagnostics of ISM properties, including the gas temperatures in the OII and OIII regions, the gas densities, metallicities, and of dust attenuation effects.\par
\begin{figure}
    \centering
    \includegraphics[width=0.45\textwidth]{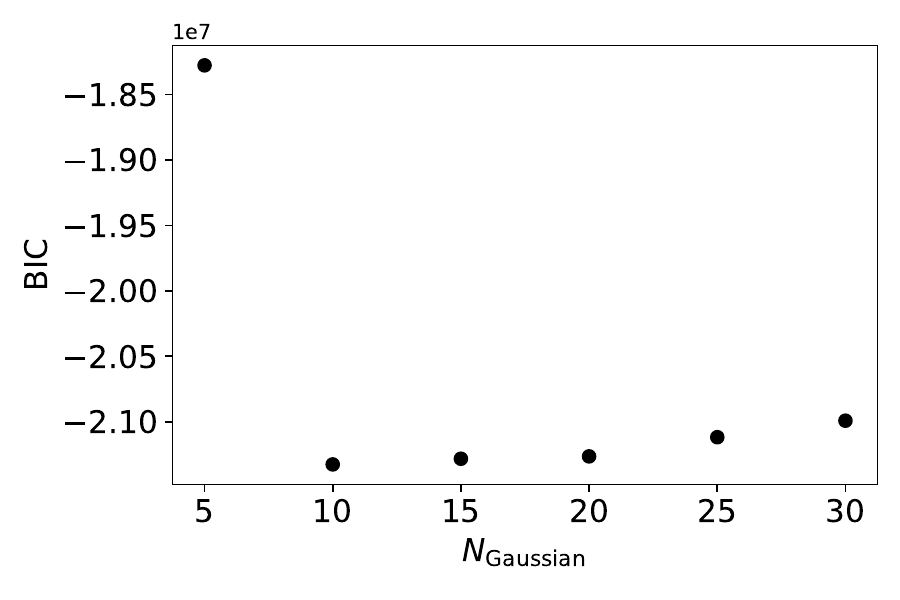}
    \caption{The BIC of our GMDN for differing numbers of Gaussian distributions in the mixture model. This suggests that $N_\mathrm{Gauss}=10$ is the optimal choice.}\label{fig:BIC}
\end{figure}

We adopt the Adam optimization algorithm with an initial learning rate of $10^{-5}$ to train the GMDN \citep{2014arXiv1412.6980K}. During the training, we feed the GMDN with FIRE data of batch size $10^5$ in each iteration. Each training is stopped after 30,000 epochs so that the loss function change after every 100 epochs of training is less than 0.01 as the training comes to an end. To determine the best value of $N_\mathrm{Gauss}$ we compute the Bayesian Information Criterion (BIC):
\begin{equation}
    \mathrm{BIC}=\mathrm{dim}\left(\theta\right)\ln n+2n\mathcal{L}\,.
\end{equation}
Here $\mathrm{dim}\left(\theta\right)$ is the number of free parameters defined in Eq~\ref{eq:dimtheta}, while $n$ is the number of HII regions in the model validation dataset. $\mathcal{L}$ is the loss function of the validation dataset for the best fit parameter $\theta$, as defined in Eq~\ref{eq:loss}. Smaller values of BIC generally correspond to more preferred models. We present the BICs under different $N_\mathrm{Gauss}$ in Figure~\ref{fig:BIC}. The BIC test suggests that the optimal choice is $N_\mathrm{Gauss}=10$.
\par

\section{Model verification}\label{sec:MDNverify}
We use the trained GMDN as an [OIII], [OII], and hydrogen recombination line emulator in the following way. Given the age and metallicity of a stellar particle, and the stellar mass of the simulated galaxy, our GMDN gives the PDF for the 4 PCA components $\phi(y)$. Since the GMM is formed by 10 4D Gaussian distributions, we first Gumbel sample to pick out one Gaussian. If the $k^\mathrm{th}$ Gaussian distribution is sampled, we pick a random number $r$ from a normal distribution $\mathcal{N}(0,1)$. The 4 PCA components are then computed as:
\begin{equation}
    \log(L_\mathrm{PCA}^i/m_*)= \sigma_k^i r+\mu_k^i\,, i=1,2,3,4\,.
\end{equation}
We then perform an inverse PCA transform to this 4D random sample to derive the 13D values of $\log(L/m_*)$. Finally, we multiply this 13D array by the stellar particle mass to derive the luminosities of the 13 modeled [OIII], [OII], H$\alpha$, and H$\beta$ lines emitted by the HII region sourced by this stellar particle. To avoid the rare case where the random sampling process hits the shallow tail of the GMM and gives unrealistically high luminosity predictions, we use the analytical model introduced in \cite{2023arXiv231209213Y} to estimate maximum values of the [OII] 3729\AA\ and [OIII] 5007\AA\ lines: 
\begin{equation}\label{eq:Lmax}
\begin{split}
    L^\mathrm{[OII]}_\mathrm{10,max}&=\left(\dfrac{n_\mathrm{O}}{n_\mathrm{H}}\right)_\odot\dfrac{Z}{Z_\odot}k_{10}^\mathrm{OII}h\nu_{10}^\mathrm{OII}\dfrac{Q_\mathrm{HI}}{\alpha_\mathrm{B,HII}}\,,\\
    L^\mathrm{[OIII]}_\mathrm{32,max}&=\left(\dfrac{n_\mathrm{O}}{n_\mathrm{H}}\right)_\odot\dfrac{Z}{Z_\odot}\dfrac{A_{32}^\mathrm{OIII}k_{32}^\mathrm{OIII}h\nu_{32}^\mathrm{OIII}}{A_{31}^\mathrm{OIII}+A_{32}^\mathrm{OIII}}\dfrac{Q_\mathrm{HI}}{\alpha_\mathrm{B,HII}}\,.
\end{split}
\end{equation}
Here to compute the upper bounds on line luminosity we assume that the gas density is much lower than the critical densities of all lines. We further assume that the OII or OIII regions span the entire HII region. In Eq~\ref{eq:Lmax}, $(n_\mathrm{O}/n_\mathrm{H})_\odot=10^{-3.31}$ is the oxygen abundance at Solar metallicity \citep{2001AIPC..598...23H}. The quantities $A_{ij}^X$, $k_{ij}^X$, and $\mu_{ij}^X$ are the spontaneous decay rate, collisional de-excitation rate, and emitted photon frequency, respectively, when particle $X$ transitions from level $i$ to level $j$. We assume case-B recombination rates throughout, and $\alpha_\mathrm{B,HII}$ denotes the case-B HII recombination rate, while $\alpha_\mathrm{B,H\beta}/\alpha_\mathrm{B,HII}$ gives the fraction of HII recombination events that lead to H$\beta$ line emission. All atomic data are temperature dependent, with values adopted from \cite{2006agna.book.....O,2011piim.book.....D}. When estimating the line luminosity upper bounds we first estimate the galaxy $Z_Q$ through the FIRE MZR at $z=6$:
\begin{equation}
    \log(Z_Q/[Z_\odot])=0.39\log(M_*/[M_\odot])-3.97\,.
\end{equation}
We then adopt metallicity-dependent OIII and OII temperature models from \cite{2023arXiv231209213Y,2020A&A...634A.107Y}. If the random sample shows a line luminosity higher than the theoretical maximum value above, we abandon this result and redo the sampling until the HII region line luminosities are all below the upper bounds. \par
\begin{figure}
    \centering
    \includegraphics[width=0.45\textwidth]{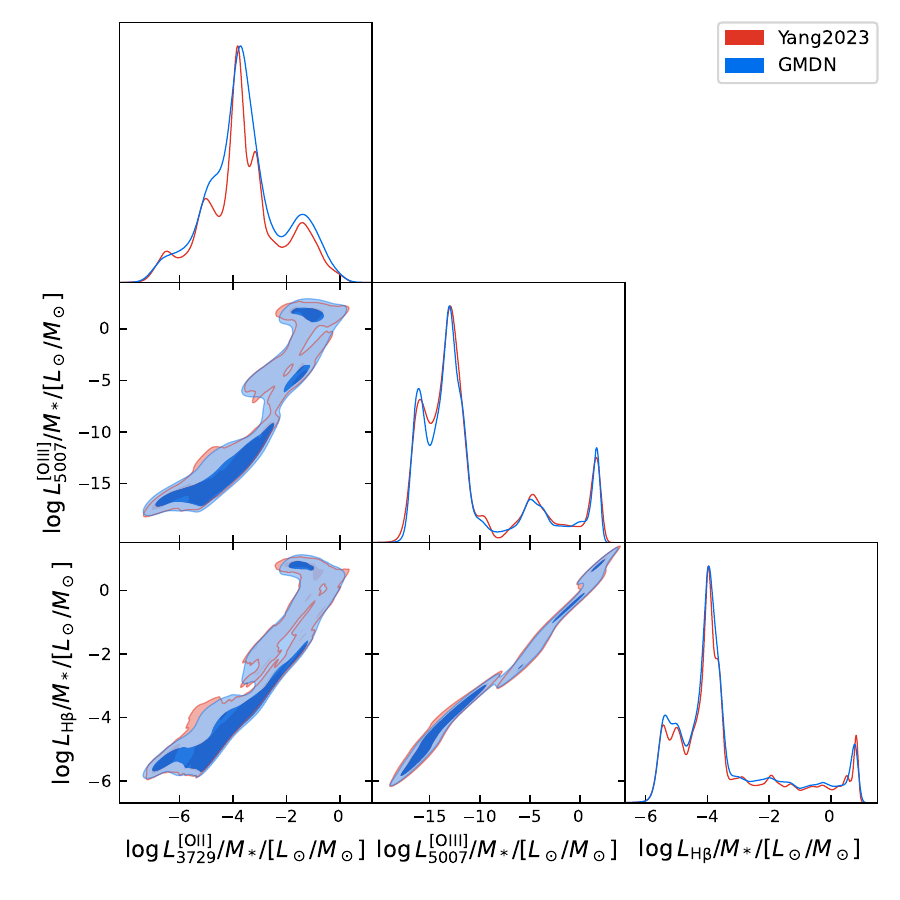}
    \caption{Luminosity distributions for the simulated and GMDN-predicted emission in the [OII] 3729 \AA, [OIII] 5007 \AA, and H$\beta$ lines for the FIRE galaxy z5m12b. The red contours show distributions from FIRE+\textsc{HIILines}, while the blue contours show the distributions of random samples generated by our GMDN. Our GMDN can accurately reproduce the distributions and correlations among [OIII], [OII], and hydrogen recombination lines.}\label{fig:triangle}
\end{figure}

In this section we verify the performance of the best fit GMDN. First, Figure~\ref{fig:triangle} compares the luminosity distributions for the [OII] 3729 \AA, [OIII] 5007 \AA, and H$\beta$ lines among all of the HII regions within the largest FIRE high-$z$ galaxy, z5m12b. The red contours show the luminosity distributions given by the FIRE simulation, while the blue contours show randomly generated data distributions from the GMDN. Figure~\ref{fig:triangle} demonstrates that the GMDN successfully reproduces the 1D distribution and 2D correlations among [OIII], [OII], and hydrogen recombination lines.
\par
\begin{figure*}
    \centering
    \includegraphics[width=0.95\textwidth]{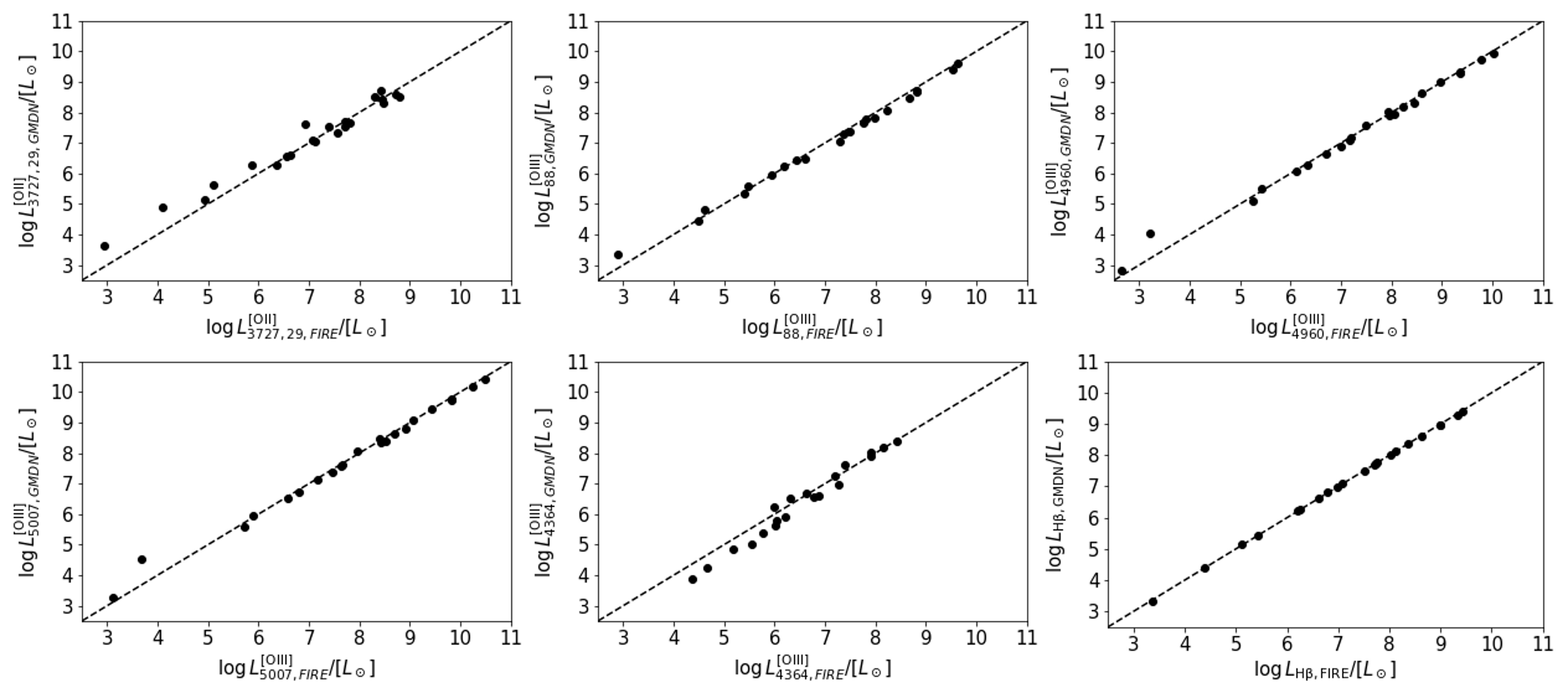}
    \caption{Galaxy-wide total [OII], [OIII], and H$\beta$ line luminosities for the 22 FIRE high-$z$ galaxies as given by \textsc{HIILines} ($x$ axis) and GMDN ($y$ axis). The black dashed lines mark equality. Each black point represents one simulated galaxy. Our GMDN successfully reproduces the galaxy-wide [OIII] 88 $\mu$m, 4960 \AA, 5007 \AA, and H$\beta$ line luminosities. It is less accurate for the fainter and more temperature sensitive [OII] 3727,29\AA\ and [OIII] 4364 \AA\ lines.}\label{fig:L_compare}
\end{figure*}

Figure~\ref{fig:L_compare} further compares the galaxy-wide total (i.e. summed over all of the HII regions in a simulated galaxy) [OII] 3729,29\AA, [OIII] 88 $\mu$m, 4960\AA, 5007\AA, 4364\AA, and H$\beta$ line luminosities given by the FIRE simulations ($x$ axis) and the GMDN ($y$ axis). 
These lines have been detected by ALMA and JWST at $z>6$, with additional observational data forthcoming in the near future. The black dashed line marks equality. The 22 FIRE galaxies used in training the GMDN are presented as black points. Figure~\ref{fig:L_compare} shows that our GMDN accurately reproduces the galaxy-wide total line luminosities. The H$\beta$ line predictions are the most accurate: the maximum $\log L$ difference between the GMDN and FIRE simulation results is only 0.04 dex. Furthermore, the GMDN performs well for lines that are weakly sensitive to gas temperature and ionization correction factor effects\footnote{In this context, the term ``ionization correction factor'' refers to our calculations of the fraction of the volume of a simulated HII region that are in OII and OIII.}, including the [OIII] 88 $\mu$m, 4960\AA, and 5007\AA\  lines. The mean of the absolute differences of $\log L$ at $\log(L/L_\odot)>4$ is 0.1, 0.07, and 0.07 dex for these three lines, respectively. The GMDN predictions for the [OII] 3727,29\AA\  and the [OIII] 4364 \AA\ auroral line are the least accurate. This is because we have used $M_*$ as an input to indirectly trace the gas temperature. However, $M_*$ and $Z_Q$ are not perfectly correlated. As a result, our GMDN method imperfectly estimates the gas temperature of simulated HII regions and is less successful when applied to temperature-sensitive lines. Overall, however, the trained GMDN accurately emulates the simulated line luminosities: it may therefore help with interpreting high-$z$ ALMA and JWST observations, and in modeling LIM signals.\par

\par
\subsection{Test of Assumptions}\label{subsec:assumptionTest} 
Our aim is to apply the trained GMDN to cosmological simulations. This involves an implicit assumption that the 22 FIRE galaxies we use to model HII region line emission, and their underlying distributions of HII region gas density and metallicity, {\em are representative. We intend to apply the FIRE-trained distributions as a sub-grid model on top of entire galaxy populations in larger volume cosmological simulations.} The potential concern here is that the ISM and line emission properties may vary more broadly than captured in our relatively small FIRE training sample.

\begin{figure*}
    \centering
    \includegraphics[width=0.95\textwidth]{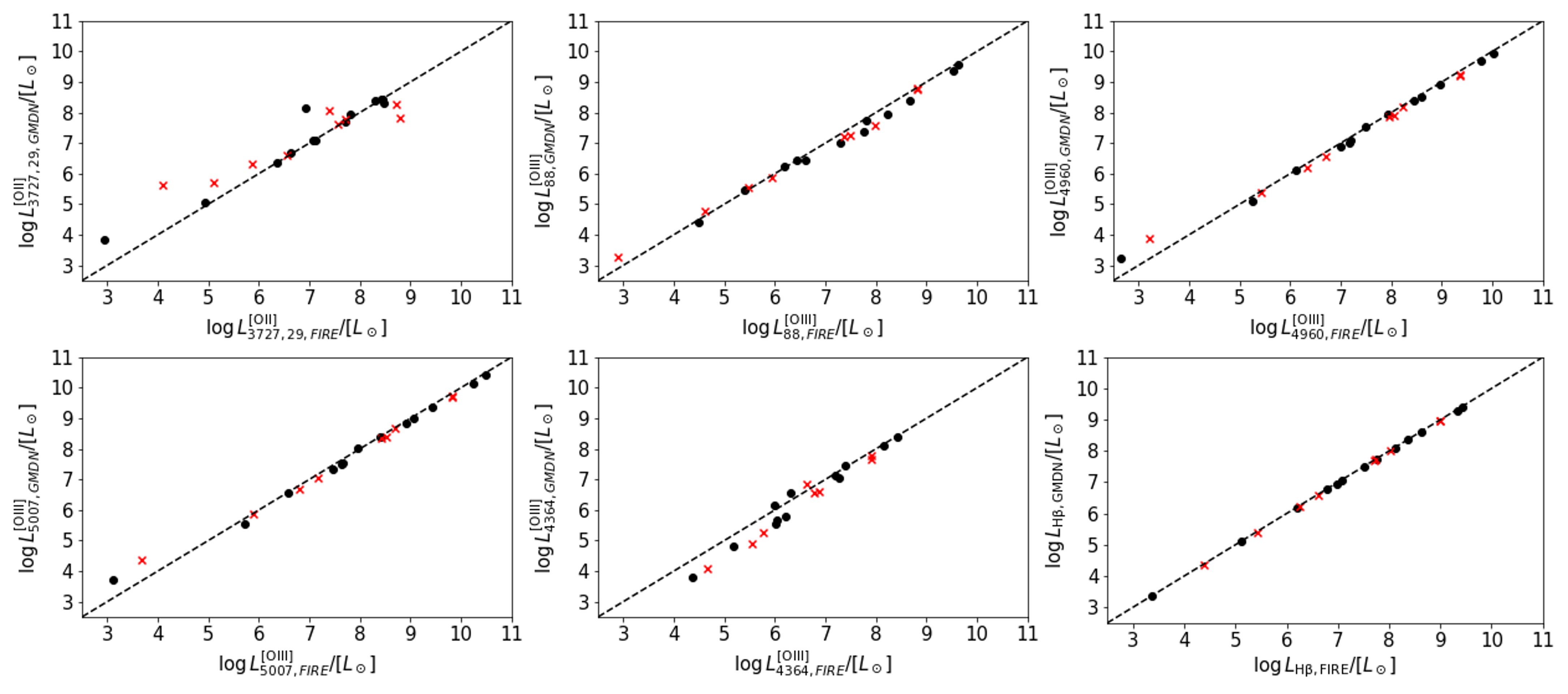}
    \caption{Same as Figure~\ref{fig:L_compare}, but here the GMDN is trained only with the 13 FIRE galaxies marked by the black points. The GMDN trained over this smaller dataset successfully reproduces the galaxy-wide [OIII], [OII], and hydrogen recombination line luminosities for the 9 galaxies that are not used in the model training (red crosses). The model performance is particularly good for the [OIII] 88 $\mu$m, 4960 \AA, 5007 \AA, and hydrogen recombination lines.}\label{fig:L_compare_test}
\end{figure*}

As one quantitative test of this assumption, we can remove galaxies from the training sample and test whether the line emission in these galaxies is well-reproduced by the sources in the new, smaller training set. The galaxies we removed from the training sample are marked with red crosses in Figure~\ref{fig:L_compare_test}. These particular sources are
excised because their stellar masses, $M_*$, are similar to those of the other FIRE galaxies which are retained in the training set. If the remaining 13 FIRE galaxies share similar ISM properties, then the GMDN should still accurately reproduce the line luminosities of the 9 sources which were excised from the training data. Indeed, Figure~\ref{fig:L_compare_test} illustrates that the reduced training data still reproduces the galaxy-wide line luminosities for the 9 excised galaxies (red crosses). As before, the performance is particularly good for the lines
that are less sensitive to gas temperature and ionization correction factor effects, including the [OIII] 88 $\mu$m, 4960 \AA, 5007 \AA, H$\alpha$, and H$\beta$ lines. The average absolute difference in the logarithm of the luminosity between
when the smaller set of galaxies and the full set are used in training is: $\log(L/[L_\odot])>4$ are 0.1, 0.1, 0.1, 0.03, and 0.03 dex for these lines, respectively.

\par

\begin{figure}
    \centering
    \includegraphics[width=0.45\textwidth]{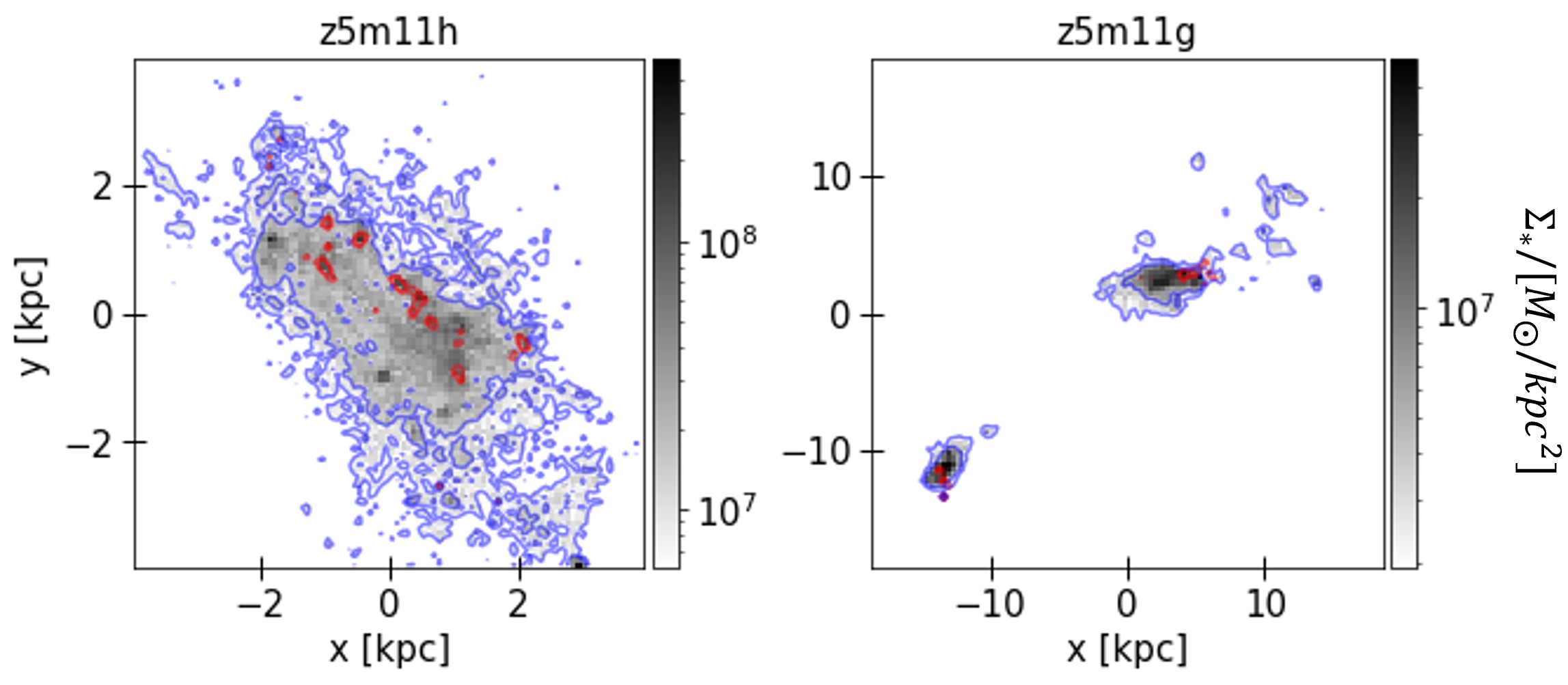}
    \caption{Face-on visualizations of the stellar mass surface density distributions for FIRE galaxies z5m11h and z5m11g. The blue and red contours enclose 50\% and 64\% of the total stellar mass and H$\beta$ line flux for each galaxy. The example galaxies, z5m11h and z5m11g, have similar stellar masses but differ strongly in their morphologies. The fact that the GMDN trained on one of these galaxies also reproduces the HII region line luminosities for the other galaxy supports our approach. The conditional line luminosity PDFs given stellar particle age, metallicity, and mass appear relatively universal, and independent of large-scale galaxy morphology and environmental properties. Note the difference in the range of scales on the x and y-axes in the examples of the left/right-hand panels. 
    }\label{fig:morphology}
\end{figure}

This test supports the notion that the conditional line luminosity PDFs involved in our analysis are relatively universal:
the FIRE sample includes galaxies with a variety of morphologies, formation histories, and environmental properties, yet
their line luminosities may be robustly captured when only a sub-set of the full sample are used in the training data. 
As an illustration, Figure~\ref{fig:morphology} shows visualizations of the FIRE galaxies z5m11h (included in our model training set) and z5m11g (excluded from the training data). 
In the visualizations, each galaxy is viewed along an axis that lies in the direction of the galaxy's total angular momentum vector. 
In each subplot, the blue (red) contours enclose 50\% and 64\% of the total galaxy stellar mass (H$\beta$ line luminosity) for each galaxy. In each panel, the grey-scale gives the stellar surface-mass density distribution. For reference, Table~\ref{tb:morphology} gives the stellar mass, H$\beta$ line luminosity, and half-mass radius for these two example galaxies. 
Although z5m11h and z5m11g have similar stellar masses, their morphologies and HII region line signal distributions differ markedly. Specifically, z5m11h is relatively compact and contains only one component, while z5m11g contains two main clumps, leading to a much larger half-mass radius. Furthermore, the galaxy-wide H$\beta$ line luminosities in these two simulated galaxies differ by a factor of 2.6. Nevertheless, our GMDN model -- which include z5m11 in its training set -- successfully reproduces the HII region line luminosities for z5m11g even when it is excised from the training data. This supports our assumption regarding the universality of the cPDFs employed, even considering galaxies across a broad range of size-scales and morphologies. 
\par

\begin{table}[]
\centering
\begin{tabular}{|c|c|c|c|}
\hline
 & $\log M_*/[M_\odot]$ & $\log L_\mathrm{H\beta}/[L_\odot]$ & $R_{50}/$[kpc]  \\ \hline
 z5m11h& 8.72  &  8.12    &   2.64          \\ \hline
 z5m11g&  8.70 &   7.70   &   9.30          \\ \hline
\end{tabular}
\caption{Stellar mass, H$\beta$ line luminosity, and half-mass radius for FIRE galaxies z5m11h and z5m11g at $z=6$.}\label{tb:morphology}
\end{table}

\section{Applying the GMDN to TNG galaxies}\label{sec:TNG}
As one example application, we apply the trained GMDN to galaxies in the publicly available IllustrisTNG simulations. IllustrisTNG is a suite of state-of-the-art cosmological galaxy formation simulations \citep{2018MNRAS.473.4077P,2018MNRAS.480.5113M,2018MNRAS.475..624N,2018MNRAS.475..676S,2018MNRAS.477.1206N,2018MNRAS.475..648P,2019MNRAS.490.3234N,2019MNRAS.490.3196P}. The publicly available IllustrisTNG simulations include cubic volumes with co-moving box lengths of 50, 100, and 300 cMpc, referred to as TNG50, TNG100, and TNG300 (with the smaller volume simulations having higher resolution). The coupled dynamics of dark matter and gas are tracked using the quasi-Lagrangian code AREPO. The simulations include models for gas heating and cooling, star formation, stellar feedback, and the formation and feedback from supermassive black holes, among other ingredients. 
The range of box size and resolution spanned by TNG50, TNG100, and TNG300 allows one to study both rare, bright galaxies and the numerous, yet smaller and less-luminous members of the galaxy population. These simulations hence provide powerful models for studying the statistical properties of galaxy populations, and this is potentially useful for comparing with JWST and ALMA emission line measurements, and near future line-intensity mapping survey data. As we have emphasized previously, however, TNG does not resolve the multi-phase structure of the ISM, let alone individual HII regions. Fortunately, our trained GMDN calculations provide FIRE-calibrated sub-grid models which can be employed to post-process realistic line emission signals on top of the stellar particles in the TNG simulations.\par 
We select galaxies at $z=6$ in TNG50, TNG100, and TNG300 that are resolved with at least 100 stellar particles. We also require that the TNG galaxies selected for post-processing contain at least one stellar particle younger than 5 Myr. This is to ensure that recent star formation activity takes place in the selected galaxies. This criterion leaves us with 7057 galaxies in TNG50, 3397 galaxies in TNG100, and 4364 galaxies in TNG300. We note that the 22 FIRE galaxies span a stellar mass range from $10^6M_\odot$ to $10^{10}M_\odot$, but 0.04\%, 0.7\%, and 3\% of TNG galaxies show stellar mass higher than this range in TNG50, TNG100, and TNG300, respectively. Since we use galaxy stellar mass as an input parameter to (indirectly) capture the HII region gas temperatures, and observational results show mild temperature variations at high metallicity/mass \citep{2023arXiv231209213Y}, we set a stellar mass upper limit of $10^{10}M_\odot$ for the GMDN when it estimates $L/M_*$ to avoid the need for any extrapolations.\par 
\subsection{Individual galaxy comparisons}
\begin{figure*}
    \centering
    \includegraphics[width=0.9\textwidth]{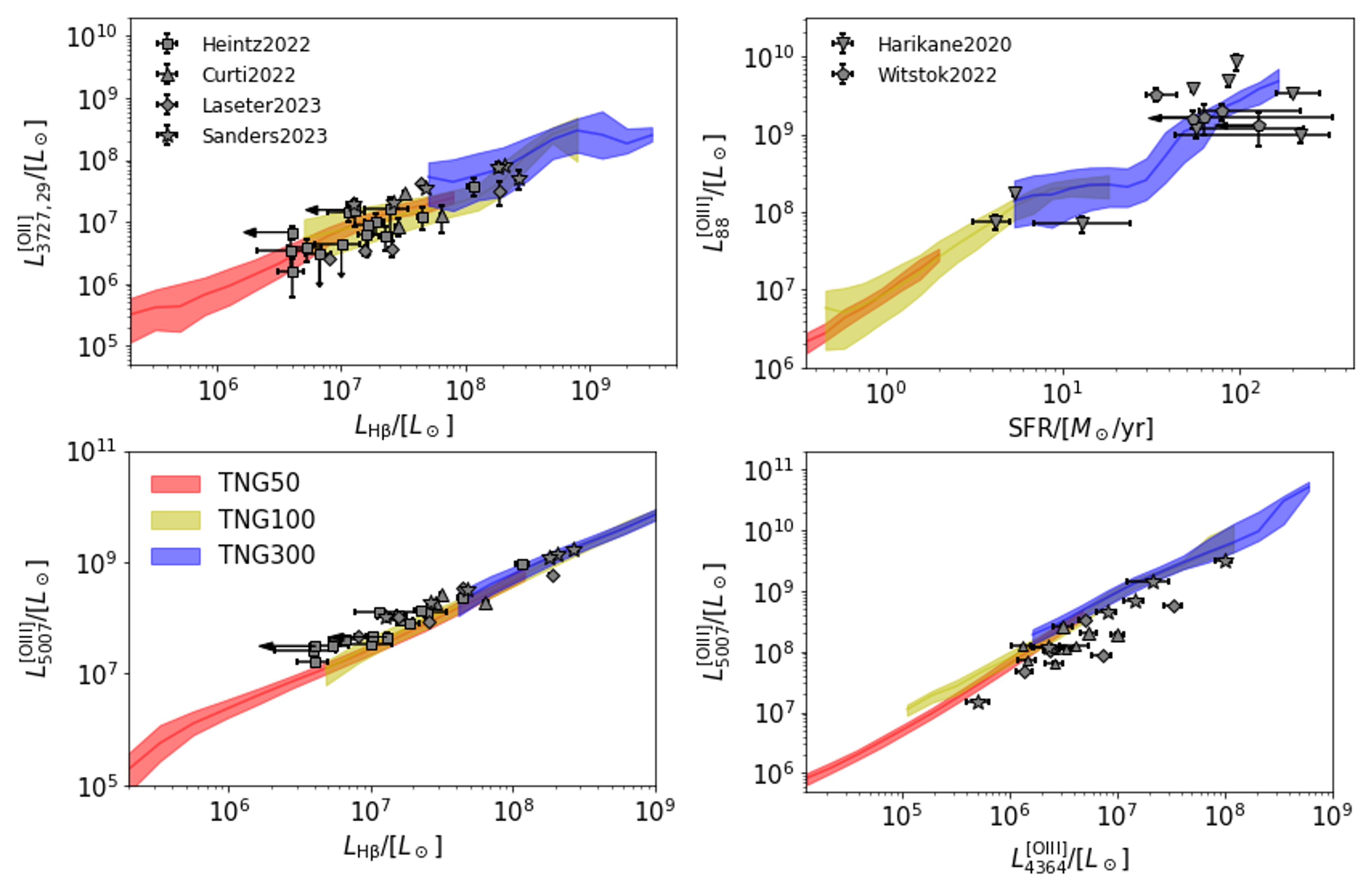}  
        \caption{Comparison of intrinsic galaxy-wide HII region line luminosities between TNG galaxies (color bands) and observations (grey points). The [OIII], [OII] rest-frame optical line signals and H$\beta$ line luminosities are from recent JWST measurements \cite{2022arXiv221202890H,2022arXiv220712375C,2023arXiv230308149S,2023arXiv230603120L}. The [OIII] 88 $\mu$m line luminosities and galaxy SFRs are from ALMA data \citep{2020ApJ...896...93H,2022MNRAS.515.1751W}. The TNG galaxies post-processed with our GMDN are generally in agreement with observations, although the GMDN under-predicts the [OIII] 4364 \AA\ auroral line luminosities by $\sim1\sigma$.}
    \label{fig:TNG_Obs}
\end{figure*}

Figure~\ref{fig:TNG_Obs} compares the post-processed ISM emission line luminosities from each TNG galaxy with current reionization-era observations  \citep{2020ApJ...896...93H,2022MNRAS.515.1751W,2022arXiv221202890H,2022arXiv220712375C,2023arXiv230308149S,2023arXiv230603120L}. Dust attenuation effects have been corrected for the JWST galaxies ID4590 and ID10612 from \cite{2022arXiv220712375C}, and all sources reported by \cite{2023arXiv230308149S,2023arXiv230603120L}, assuming an extinction curve with $R_V=2.5$ \citep{2011piim.book.....D}.
The other JWST galaxies in the figure are measured to have negligibly small dust extinctions with $A_V < 0.25$ mag (and most such sources have less than 0.1 magnitude of extinction).
The observational results (grey crosses) generally agree well with the TNG results, as shown in the red, yellow, and blue bands which indicate the line luminosities post-processed on top of the TNG50, TNG100, and TNG300 simulations, respectively. The bands encompass the 16\% to 84\% range in simulated line luminosities, while the solid lines indicate the median line luminosities. 
However, the bottom right panel of Figure~\ref{fig:TNG_Obs} shows that our method underestimates the [OIII] 4364\AA\ line luminosities by $\sim1\sigma$. This is because, at a fixed SFR, the 22 FIRE galaxies tend to show higher metallicities and lower HII region temperatures than the measured galaxies, and this leads to underestimates of the temperature-sensitive [OIII] 4364\AA\ line luminosities. In summary, our trained GMDN, which uses FIRE to develop an ISM line emission sub-grid model for TNG, matches current
observations fairly well. The main shortcoming is in simulating highly gas temperature-sensitive lines, where our current method appears less accurate. \par

\subsection{$L(\mathrm{SFR})$ relations}
In general, the line luminosities from our simulated galaxies are well-correlated with their SFRs: galaxies with high SFRs produce copious numbers of ionizing photons (via their young and hot O and B stars), which source HII regions and the lines of interest for our study. Of course the trend of line luminosity with SFR and the level of scatter around the mean relation vary from line to line.
Here we explore these trends using our simulated TNG line-emitting galaxies, and compare with earlier work in the literature.
Ultimately, one related goal here is to use LIM observations to extract the luminosity density as a function of redshift in some of these lines (e.g. \cite{2017ApJ...835..273G,2021ApJ...915...33S,2022A&ARv..30....5B}). Provided the $L(\mathrm{SFR})$ relation is understood or empirically-calibrated, these observations can then be used to determine the star formation rate density as a function of redshift, including contributions from low luminosity sources which lie beneath the flux limits of traditional surveys. 

\begin{figure*}
    \centering
    \includegraphics[width=0.95\textwidth]{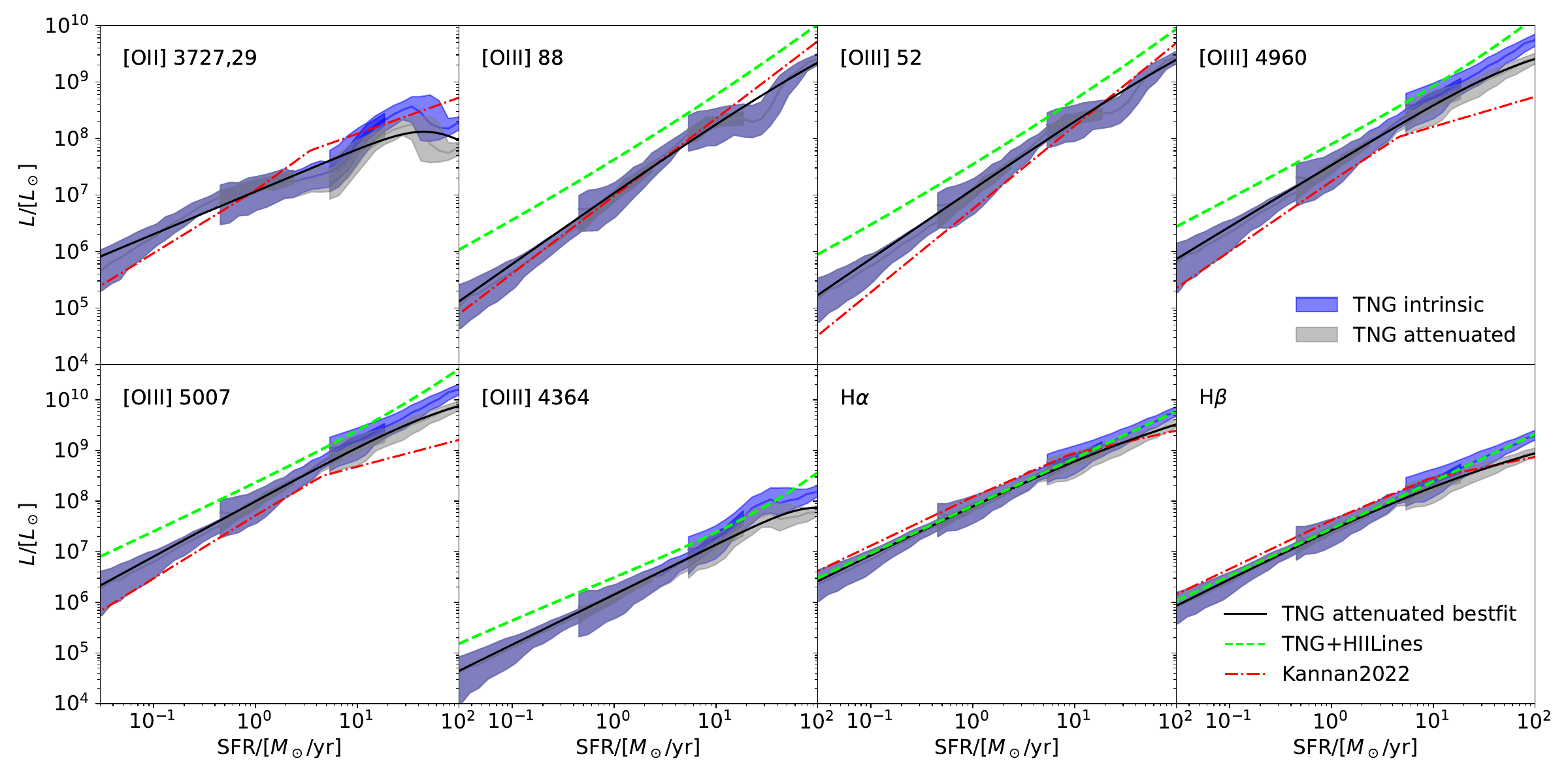}  
        \caption{Line luminosity versus SFR relations given by TNG50, TNG100, and TNG300 in our models, compared to simulation results from \cite{2022MNRAS.514.3857K} (red dash dotted curves). The blue bands show the median and 1$\sigma$ scatter of our intrinsic (i.e. without dust extinction) GMDN $L(\mathrm{SFR})$ relations, while the grey bands include dust attenuation estimates at the relevant wavelengths. The best fit double power-law relations from our dust-attenuated GMDN $L(\mathrm{SFR})$ models are shown as solid black curves. The lime dashed curves show $L(\mathrm{SFR})$ models based on applying \textsc{HIILines} directly on top of TNG, without including the FIRE-calibrated sub-grid models.   
        }
    \label{fig:LSFR}
\end{figure*}

Figure~\ref{fig:LSFR} compares the line luminosity versus SFR relation predicted by TNG$+$GMDN and \cite{2022MNRAS.514.3857K} (hereafter \defcitealias{2022MNRAS.514.3857K}{Kannan2022}\citetalias{2022MNRAS.514.3857K}). The blue bands shows the median and 1$\sigma$ scatter in the intrinsic line luminosities predicted by our method, while the grey bands show post-processed line signals after accounting for dust attenuation at the wavelengths of interest for each emission line. In order to model dust attenuation we use the observationally-motivated dust Model A from \cite{2020MNRAS.492.5167V}, which takes the absolute UV magnitude of each simulated galaxy as input. 
We assume an extinction curve from \cite{1989ApJ...345..245C} with an average Milky Way extinction per reddening of $R_V=3.1$. On the other hand, \defcitealias{2022MNRAS.514.3857K}{Kannan2022}\citetalias{2022MNRAS.514.3857K} applies the more sophisticated and expensive Model C of \cite{2020MNRAS.492.5167V}, where a Monte Carlo radiative transfer code is used to model the dust attenuation.\par 
Here we want to re-emphasize the main difference between this work and most other methods that post-process ISM emission lines on top of cosmological simulations. In most earlier work, including \defcitealias{2022MNRAS.514.3857K}{Kannan2022}\citetalias{2022MNRAS.514.3857K}, the authors either directly adopt the often poorly resolved ISM properties in the simulations employed, or adopt simplifications such as assuming constant density HII regions and/or incident radiation fields.
These simplifications were necessary in part because of the limited resolution of the simulations and partly because the line emission calculations are generally performed using \textsc{Cloudy} or similar codes. The \textsc{Cloudy} calculations become extremely expensive when each stellar or gas particle in the simulation must be modeled separately, and so this forces simplifications such as assuming that all emitting HII regions have identical gas densities. 
Our approach refines the earlier line signal modeling efforts through introducing a sub-grid model to capture the small-scale ISM properties which are not directly resolved in the large volume simulations. 
Specifically, our GDMN allows us to produce random samples of the line emission around stellar particles in a larger volume cosmological simulation, with the statistical properties of this line emission matching those of the 22 FIRE zoom-in galaxies. The line emission from the FIRE galaxies is in good agreement with current $z \gtrsim 6$ line luminosity observations.
Additionally, the computationally efficient \textsc{HIILines} allows the FIRE galaxy line emission calculations to fully account for variations in the stellar radiation spectral shape, amplitude, ISM gas density, and metallicity. As discussed previously, this is in contrast to previous \textsc{Cloudy}-based analyses which generally require simplifying assumptions (such as assuming a constant gas density and radiation spectrum across all HII regions), owing to their computational expense. \par
We fit the mean line luminosity (while including a suppression factor to account for dust attenuation at the relevant wavelengths for each line) versus SFR relations given by the GMDN with a double power law \citep{2018MNRAS.475.1477P,2022ApJ...929..140Y}:
\begin{widetext}
\begin{equation}\label{eq:dplaw}
    \dfrac{L}{[L_\odot]}=2N\dfrac{\mathrm{SFR}}{[M_\odot/\mathrm{yr}]}\left[\left(\dfrac{\mathrm{SFR/[M_\odot/\mathrm{yr}]}}{\mathrm{SFR}_1}\right)^{-\alpha}+\left(\dfrac{\mathrm{SFR}/[M_\odot/\mathrm{yr}]}{\mathrm{SFR}_1}\right)^\beta\right]^{-1}\,.
\end{equation}
\end{widetext}
The best fit parameters are summarized in Table~\ref{tb:LSFR}. There are several interesting differences between the best fit relations from this work (black solid curves) and \defcitealias{2022MNRAS.514.3857K}{Kannan2022}\citetalias{2022MNRAS.514.3857K} (red dash-dotted curves) in Figure~\ref{fig:LSFR}. In \defcitealias{2022MNRAS.514.3857K}{Kannan2022}\citetalias{2022MNRAS.514.3857K}, all of the $L(\mathrm{SFR})$ relations for the optical lines show shallower slopes at high SFR. This is caused by the stronger dust attenuation effect in metal enriched galaxies. The $L(\mathrm{SFR})$ slopes at high SFR are generally different between these two works. This is likely caused by our different dust model assumptions. In our work,  $L_\mathrm{[OII]}(\mathrm{SFR})$ also bends downward at high SFR partly because the high SFR galaxies tend to contain younger stellar populations which emit harder ionizing radiation spectra. This, in turn, increases the fractional OIII volume while it decreases the OII volume and [OII] luminosity within the HII regions in these galaxies.  
Furthermore, the high SFR galaxies also generally have higher gas densities, which also leads to larger OIII fractional volumes \citep{2023MNRAS.525.5989Y}.
In the case of the sub-mm [OIII] lines, dust attenuation is negligible, and so \defcitealias{2022MNRAS.514.3857K}{Kannan2022}\citetalias{2022MNRAS.514.3857K} adopt
linear fits for $L_\mathrm{[OIII]88}(\mathrm{SFR})$ and $L_\mathrm{[OIII]52}(\mathrm{SFR})$.
However, in our models $L_\mathrm{[OIII]88}(\mathrm{SFR})$ and $L_\mathrm{[OIII]52}(\mathrm{SFR})$ still show small suppressions (relative to linear relations) at high SFR. In our calculations this occurs again because high SFR galaxies tend to be denser. The higher collisional de-excitation rates at high density lead to transitions without photon emission and a reduction in the line luminosity to SFR ratio \citep{2020MNRAS.499.3417Y}. This effect is missed in \citetalias{2022MNRAS.514.3857K} because the ISM gas density is fixed in that work to $n_e$ =100 cm$^{-3}$, and this is lower than the critical densities of these lines.
However, in our FIRE-calibrated sub-grid model there are numerous HII regions at higher densities, especially within high SFR galaxies. Finally, note that our calculations assume sub-grid metallicities from FIRE, while \defcitealias{2022MNRAS.514.3857K}{Kannan2022}\citetalias{2022MNRAS.514.3857K} adopt metallicites directly from the TNG simulations. This accounts for some of the differences between our models, as the FIRE and TNG MZRs are disparate (see Figure~\ref{fig:MZR}). 
\begin{table}[]
\centering
\begin{tabular}{|c|c|c|c|c|}
\hline
 Line & N & SFR$_1$ & $\alpha$ & $\beta$ \\ \hline
[OII] 3727,29 &2.00E6   & 6.28E1     & -2.53E-1      & 2.49     \\ \hline
[OIII] 88& 2.20E7  & 1.57E2     &  2.70E-1     &  3.24E-1    \\ \hline
[OIII] 52& 2.34E7  & 2.71E2     &  2.33E-1     & 4.72E-1     \\ \hline
[OIII] 4960&2.61E7   &  9.60E1    &  9.46E-2     & 9.24E-1     \\ \hline
[OIII] 5007& 7.68E7  & 9.76E1     & 9.48E-2      &  9.28E-1    \\ \hline
[OIII] 4364&6.78E5   & 1.09E2     & -9.62E-3      &  2.25    \\ \hline
H$\alpha$& 4.28E7  & 4.42E1     &  -4.03E-3     &  5.84E-1    \\ \hline
H$\beta$& 1.63E7  &  1.79E1    & 1.78E-2      &   5.77E-1   \\ \hline
\end{tabular}
\caption{Best fit parameters for the dust attenuated $L(\mathrm{SFR})$ relations parameterized by Eq~\ref{eq:dplaw}.}\label{tb:LSFR}
\end{table}

\begin{figure}
    \centering
    \includegraphics[width=0.45\textwidth]{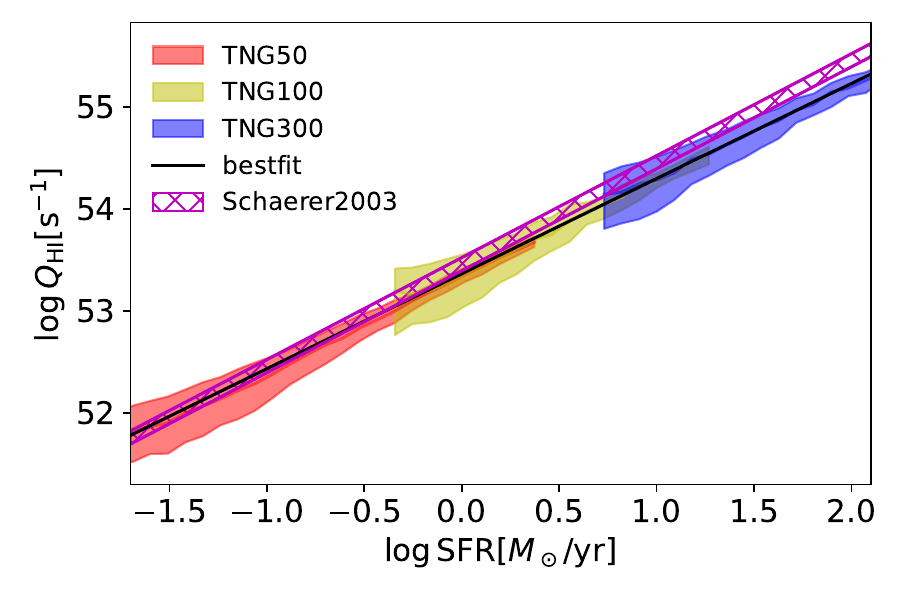} \\ 
    \includegraphics[width=0.45\textwidth]{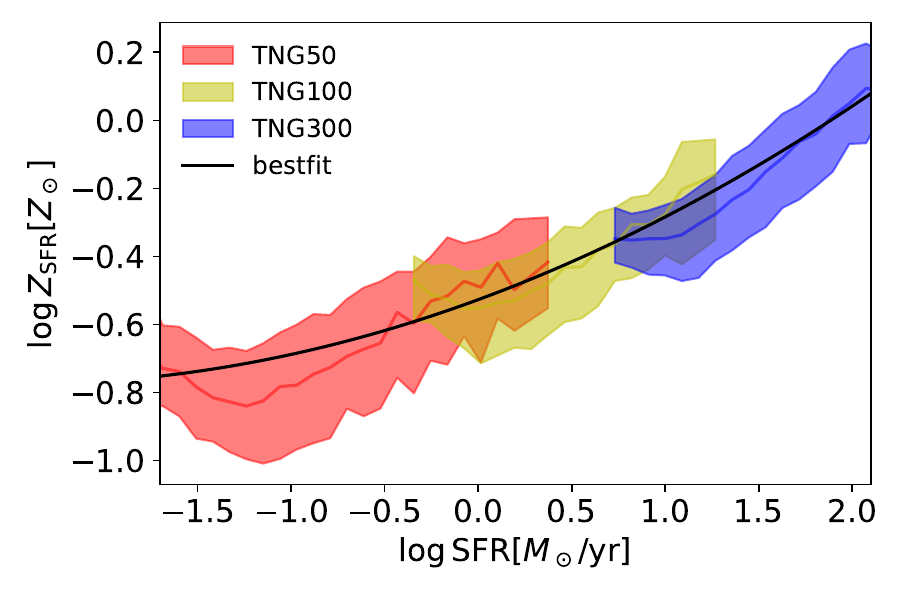} 
        \caption{Correlation between each of the galaxy-wide hydrogen ionizing photon generation rate, $Q_\mathrm{HI}$ (top panel), and the 
        SFR-weighted metallicity,  $Z_\mathrm{SFR}$ (bottom panel), and SFR. 
        The red, yellow, and blue lines show the median relations for the TNG50, TNG100, and TNG300 simulations, respectively, while the bands enclose the relations for 16\% and 84\% of the simulated galaxies. 
        The best fits for the average $Q_\mathrm{HI}(\mathrm{SFR})$ and $Z_\mathrm{SFR}(\mathrm{SFR})$ relations are shown by the black curves. For comparison, $Q_\mathrm{HI}(\mathrm{SFR})$ from the fitting formula of \protect\cite{2003A&A...397..527S}, within the stellar metallicity range of $0.1\leq Z/[Z_\odot]\leq1$, is shown by the magenta band.}\label{fig:TNGISM}
\end{figure}

We can obtain a more quantitative understanding of the differences between the ISM properties in FIRE and TNG, and how these influence the $L(\mathrm{SFR})$ relations. 
First, we can directly adopt the TNG ISM properties and apply \textsc{HIILines} to predict $L(\mathrm{SFR})$ without including our FIRE-calibrated sub-grid model. 
These calculations should resemble previous work in the literature, including \defcitealias{2022MNRAS.514.3857K}
{Kannan2022}\citetalias{2022MNRAS.514.3857K}. 
In order to explore this, we must first determine some relationships between the rate of ionizing photon production and SFR among TNG galaxies, and their gas phase metallicities.
First, the top panel of Figure~\ref{fig:TNGISM} shows the median TNG galaxy $Q_\mathrm{HI}$ versus SFR relation, and its 1$\sigma$ scatter. For each TNG galaxy, we derive $Q_\mathrm{HI}$ by summing over all stellar particles younger than 100 Myr in age, after
assuming an FSPS stellar population synthesis model and a Chabrier initial mass function (IMF). The resulting best-fit mean
$Q_\mathrm{HI}(\mathrm{SFR})$ relation (black curve) is given by:
\begin{equation}
    \log\left(\dfrac{Q_\mathrm{HI}}{[s^{-1}]}\right)=0.932\log\left(\dfrac{\mathrm{SFR}}{[M_\odot/\mathrm{yr}]}\right)+53.4\,.
\end{equation}
For reference, we note that the average TNG $Q_\mathrm{HI}(\mathrm{SFR})$ relationship is close to the fitting formula from \cite{2003A&A...397..527S} for a constant SFR model with a Salpeter IMF and stellar population ages larger than 6 Myr. (See the magenta band, where the upper and lower bounds adopt constant stellar metallicities between 0.1 and 1 $Z_\odot$, respectively.) 
Further, the average SFR-weighted gas-phase metallicity, which serves to mimic the $Q_\mathrm{HI}$-weighted metallicity, for all TNG galaxies is shown in the bottom panel of Figure~\ref{fig:TNGISM}. The best fit result is:
\begin{equation}
\begin{split}
    \log\left(\dfrac{Z_\mathrm{SFR}}{[Z_\odot]}\right)&=0.0411\log\left(\dfrac{\mathrm{SFR}}{[M_\odot/\mathrm{yr}]}\right)^2\\
    &+0.202\log\left(\dfrac{\mathrm{SFR}}{[M_\odot/\mathrm{yr}]}\right)-0.527\,.
\end{split}
\end{equation}
We can then estimate the HII and OIII region gas temperatures with the empirically-motivated relation \citep{2023arXiv231209213Y}:
\begin{equation}\label{eq:TISM}
    T_4=0.824(\log Z_\mathrm{SFR})^2+0.101\log Z_\mathrm{SFR}+1.08\,.
\end{equation}
Here $Z_\mathrm{SFR}$ is in units of $Z_\odot$, and $T_4$ is the ionized ISM gas temperature in units of $10^4$K.\par 

Using these relationships as input, and further assuming a constant ISM gas density of $n_{\mathrm e} = 100$ cm$^{-3}$ and that the volume of each OIII region matches the HII region volume, $V_\mathrm{OIII}/V_\mathrm{HII}=1$, 
we compute the intrinsic $L(\mathrm{SFR})$ relations for all of the [OIII] and hydrogen recombination lines from \textsc{HIILines}.
The TNG$+$HIILines results, which ignore our FIRE sub-grid modeling steps, are shown by the lime dashed curves in Figure~\ref{fig:LSFR}. 
First, note that the GMDN predictions for the H$\alpha$ and H$\beta$ luminosities overlap strongly with the simplified TNG$+$HIILines
models. This occurs because the hydrogen recombination lines are fairly insensitive to local ISM properties, such as the HII region gas density and metallicity. Instead, the luminosities in the Balmer lines are mostly determined by the $Q_\mathrm{HI}$ of each galaxy, and this is well-captured by the stellar particle information from TNG alone (and the stellar population synthesis model).
Nevertheless, the \defcitealias{2022MNRAS.514.3857K}{Kannan2022}\citetalias{2022MNRAS.514.3857K} $L(\mathrm{SFR})$ relations for these two lines are slightly higher than our predictions. This might be caused by the fact that we have assumed a different IMF for the simulated stellar populations. 
On the other hand, the TNG$+$HIILines $L(\mathrm{SFR})$ relations for the [OIII] 88 and 52 micron lines are significantly higher than the GMDN results.
This mainly results because the former method adopts the TNG gas-phase metallicities directly, while the GMDN results are based on the FIRE simulation metallicities. Figure~\ref{fig:MZR} shows that the TNG metallicities are higher than those in FIRE by as much as $\sim 1$ order of magnitude, leading to the discrepant line luminosity predictions. 
Finally, the TNG$+$HIILines $L(\mathrm{SFR})$ relations for the [OIII] rest-frame optical lines are also higher than the GMDN post-processing results, although the difference is less drastic compared with the [OIII] sub-millimeter lines. This is because, with the higher TNG gas-phase metallicities, Eq~\ref{eq:TISM} predicts lower ISM temperatures for each galaxy, and this suppresses the [OIII] optical line luminosities. We have not yet, however, found an explanation as to why the \defcitealias{2022MNRAS.514.3857K}{Kannan2022}\citetalias{2022MNRAS.514.3857K} [OIII] $L(\mathrm{SFR})$ results differ from our TNG$+$HIILines predictions even at low SFR. \par

\subsection{Luminosity Functions}

The line luminosity functions, $\phi(L)$, encode interesting information regarding ISM properties, stellar radiation fields, and also
the population-level statistics of the emitting galaxies. More specifically, in the case of the Balmer lines, the line luminosity functions directly trace the overall ionizing emissivity of the emitting galaixes. Assuming photo-ionization equilibrium and case-B
recombinations (which depend weakly on temperature), the rate of ionizing photon production (asides for those that escape the galaxy to ionize atoms in the IGM) may be translated directly into a Balmer line luminosity. Hence, after accounting for dust attenuation and the likely small average escape fraction of ionizing photons, the Balmer line luminosity function can be used to infer the rate of ionizing photon production per unit volume. This ionizing emissivity is a key quantity for understanding reionization, and so the Balmer line luminosity functions should provide important empirical guidance here. The [OIII] and [OII] line luminosity functions yield
information regarding the chemical enrichment history of the universe, and the properties of the ISM, especially when compared to the Balmer line luminosity functions. Furthermore, the relative luminosity functions of [OIII] and [OII] depend on the typical shape of the ionizing radiation spectrum and should thereby provide insights into the statistical properties of the stellar populations in the emitting galaxies. Furthermore, current and upcoming measurements, including new JWST observations, will deliver on the scientific promise here. In addition, line-intensity mapping surveys effectively probe integrals over the luminosity functions, including the impact of low-luminosity sources (e.g. \citealt{2022A&ARv..30....5B}). 

\begin{figure*}
    \centering
    \includegraphics[width=0.95\textwidth]{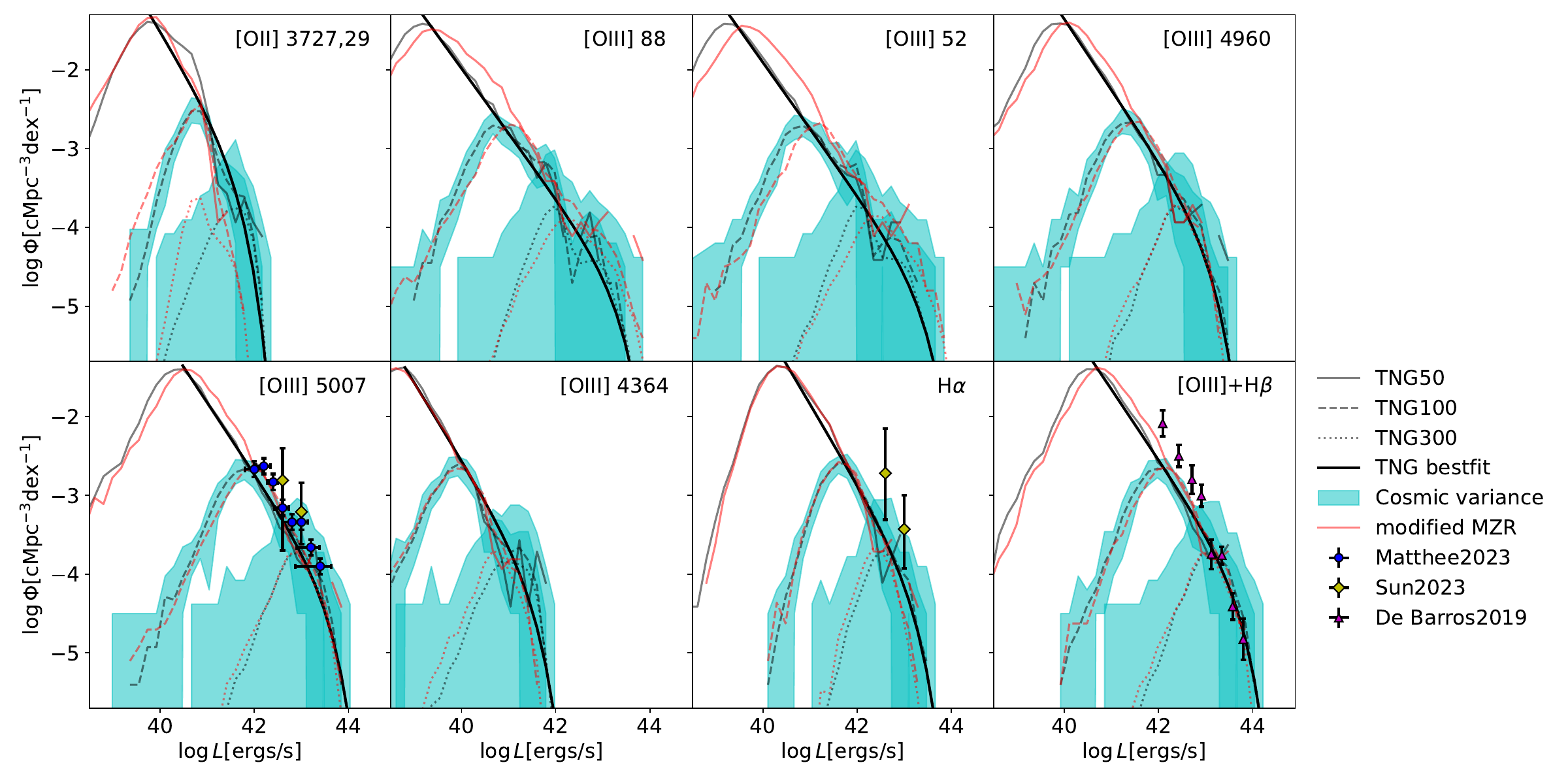}  
        \caption{Comparisons between our GMDN model luminosity functions (black transparent curves) and current observations \citep{2019MNRAS.489.2355D,2023ApJ...950...67M,2023ApJ...953...53S}. Schechter function fits to the simulation results are shown as solid black curves, and the corresponding best-fit parameters are summarized in  Table~\ref{tb:LF}. The red transparent curves show luminosity function models in which the FIRE galaxy stellar and gas particle metallicities are boosted by 0.4 dex. This model is intended to mimic the TNG MZR, in contrast to our fiducial case where the MZR matches that from the FIRE simulations.
        }
    \label{fig:LF}
\end{figure*}

Motivated by these exciting prospects, the black curves in Figure~\ref{fig:LF} show $z=6$ line luminosity function results from our new GMDN models on top of TNG50 (solid), TNG100 (dashed), and TNG300 (dotted). The combined luminosity functions are fit to Schechter function models:
\begin{equation}\label{eq:LF}
    \phi(L)=\dfrac{dn}{dL}=\phi^*\left(\dfrac{L}{L^*}\right)^\alpha\exp(-(L/L^*))\dfrac{1}{L^*}\,,
\end{equation}
where $dn$ denotes the co-moving number density of galaxies in a given line luminosity bin, i.e. with line luminosity between $L$ and $L + dL$. The best-fit results are given by solid curves in Figure~\ref{fig:LF} and the best-fit parameters are summarized in Table~\ref{tb:LF}.\par 

We also compare the simulated luminosity functions with observations from \cite{2019MNRAS.489.2355D,2023ApJ...950...67M,2023ApJ...953...53S}. Our simulated [OIII] 5007 \AA\ luminosity function model agrees with observational results to within their $1\sigma$ uncertainties in most luminosity bins, although the model lies consistently beneath the observational data.
Likewise, the simulated $H\alpha$ and $\mathrm{[OIII] 4960, 5007 + H\beta}$ luminosity functions fall below the current observational results by more than $1\sigma$. In fact, previous studies (without our sub-grid modeling framework) generally found
more pronounced discrepancies between these measurements and models (e.g. \citealt{2023MNRAS.526.3610H}). 

One possibility is that part of the discrepancies between luminosity function model and measurements owes to sample/cosmic variance.
That is, the measurements are performed over small regions of the sky and so the cosmic variance uncertainties can be large, yet these are not generally incorporated into the observational error budgets. If the observed fields happen to be centered near
over-dense regions, they could yield larger luminosity function estimates than expected based on our ensemble-averaged models.  
For example, the H$\beta$+[OIII] luminosity function measurement comes from the EIGER survey \citep{2023ApJ...950...67M}. This survey spans 25.9 arcmin$^2$ over the redshift range from $5.33<z<6.96$, corresponding to a volume of $1.06\times10^5$ cMpc$^3$. 
This is similar to the volume spanned by the TNG50 simulation box. In order to assess the level of cosmic variance in the EIGER luminosity function measurements, we hence divide the TNG100 and TNG300 simulation boxes into 8 and 216 cubical sub-volumes to roughly
match the EIGER survey volume. The spread in the luminosity function estimates across these sub-volumes then provides an estimate of the cosmic variance contributions to the luminosity function uncertainties.\footnote{Note that to calculate the cosmic variance we assume cubical regions which match the EIGER survey in volume. A more accurate cosmic variance estimate would also mimic the rectangular box geometry of the survey. Properly matching the survey geometry here would boost the cosmic variance uncertainties \citep{2010MNRAS.407.2131D}, and so our cubical treatment likely provides conservative underestimates of the error budget here.} The minimum to maximum luminosity function ranges are shown as cyan bands in Figure~\ref{fig:LF}. As can be discerned from the figure, the cosmic variance contributions to the error budget are important and can explain away much of the discrepancies with our models. Larger volume surveys will be required to enable stronger tests of these models in the future.\par

A virtue of our modeling framework is that it is efficient and flexible enough to allow us to explore variations around the input assumptions and thereby gain insights into which features of the models are most important. In order to obtain a better understanding of how the FIRE and TNG simulation properties influence our line luminosity function model predictions, we recalculate the luminosity functions after varying the assumed MZR model. As illustrated in Figure~\ref{fig:MZR}, the TNG MZR shows a higher normalization (i.e. larger metallicities at a given stellar mass) than in FIRE by about $\sim 0.4$ dex. We hence explore the impact of manually boosting the metallicities among both the stellar and gas particles in FIRE by 0.4 dex. We then model the line emission from the modified FIRE HII regions, re-train the GMDN, and use this to predict modified post-processed emission line signals from the TNG simulation. The resulting modified luminosity function models
are shown as red curves in Figure~\ref{fig:LF}. As expected, the [OIII] 88 and 52 $\mu$m luminosity functions are shifted towards
higher luminosities by a factor of $\sim 2.5$ owing to the enhanced metallicities in the case of the modified MZR model. The luminosity functions in the rest-frame optical [OIII] lines also show luminosity enhancements, but the shifts are smaller in these lines. This occurs, because while increasing the metallicity of each galaxy boosts its luminosity, there is a counteracting decrease since the more metal rich gas leads to lower gas temperatures in our models. Finally, the Balmer lines are insensitive to variations in the MZR model as hydrogen lines are metallicity independent, and the only small effect arises through the metallicity-dependent shift in the gas temperature, which weakly impacts the Balmer line strengths.\par 

Other factors that can influence the simulated luminosity functions include switching from the FSPS single star model to binary star radiation spectra, or adopting different dust attenuation treatments. We will explore these possibilities quantitatively in future work. Overall, it will be instructive to compare the upcoming HII region line luminosity function measurements with the simulation predictions: discrepancies should help refine the simulation models. 

\begin{table}[]
\begin{tabular}{|c|c|c|c|}
\hline
 Line & $\log(\phi^*)$ & $\log(L^*)$ & $\alpha$  \\ \hline
[OII] 3727,29 & -3.35  & 41.6    &  -1.99         \\ \hline
[OIII] 88& -5.08  &   43.3   &  -1.82         \\ \hline
[OIII] 52& -5.12  &  43.4    &  -1.84         \\ \hline
[OIII] 4960& -4.47  &  43.1    &  -1.90         \\ \hline
[OIII] 5007& -4.47  & 43.5     &  -1.90       \\ \hline
[OIII] 4364& -4.29  &  41.5    & -1.95         \\ \hline
H$\alpha$& -4.39  &  43.2    &   -2.01        \\ \hline
H$\beta$& -4.23  & 42.5     &  -2.03       \\ \hline
[OIII]$+$H$\beta$& -4.40  & 43.7     & -1.89        \\ \hline
\end{tabular}
\caption{Best fit parameters for the parameterized luminosity function Eq~\ref{eq:LF}. The unit of $\phi^*$ is $[\mathrm{cMpc\ ergs^{-1} s}]$. The unit of $L^*$ is ergs/s. $\mathrm{[OIII]+H}\beta$ stands for the total luminosity of [OIII]4960, 5007, and the H$\beta$ lines.}\label{tb:LF}
\end{table}

\section{Summary and discussion}\label{sec:summary}
Current and forthcoming emission line measurements from the JWST, ALMA, and SPHEREx allow new tests of state-of-the-art simulations of galaxy formation. This requires, however, producing detailed models of ISM emission lines across cosmological volumes.  
A difficulty here is that the line emission is shaped by the multi-phase properties of the ISM, yet this small-scale structure remains unresolved in large-volume cosmological simulations.
To overcome this challenge, we introduced a novel, multi-scale modeling framework which combines state-of-the-art zoom-in galaxy simulations, a semi-analytic HII region emission line model, and the MDN machine learning architecture.\par 
In a previous work, \cite{2023MNRAS.525.5989Y}, we applied a spectral synthesis model, \textsc{HIILines}, to 22 FIRE central galaxies at $z=6$ to model the [OIII], [OII], and hydrogen recombination line luminosities among all of the individual HII regions across each simulated galaxy. 
The FIRE simulations provide self-consistent galaxy-scale models at $\sim 10$ pc resolution, partly capturing the properties of the multi-phase ISM. 
In addition, these 22 FIRE galaxies span a wide range in mass, metallicity, and morphology, and are therefore representative of a wide variety of the galaxies found in larger volume cosmological simulations. The strength of \textsc{HIILines} is its high computational efficiency: this allows it to account for variations in the stellar particle and local ISM properties, avoiding simplified treatments where the gas density and/or radiation field are assumed constant across all HII regions and galaxies.  
The galaxy-wide [OIII], [OII], hydrogen recombination line luminosities, line ratios, luminosity versus SFR relations, MZRs, and UV luminosity functions given by the FIRE high-$z$ simulations agree well with current measurements \citep{2023MNRAS.525.5989Y,2023ApJ...955L..35S}. The FIRE simulations hence serve as the basis for our sub-grid line emission models.\par
Here we used the 22 FIRE galaxies as a training set and the GMDN technique to extract joint cPDFs for the [OIII] [OII], 
H$\alpha$, and H$\beta$ line luminosities per stellar particle mass, conditioned on the stellar particle age, metallicity, and total galaxy stellar mass.
In this trained GMDN model, the local ISM properties among the FIRE galaxies are effectively folded into the free parameters in the network. Assuming that independent sub-grid HII regions exist around each stellar particle in a larger volume cosmological simulation or semi-analytic model, the GMDN acts as a generative model to populate such sub-grid HII regions with line luminosities. It can be applied to any simulation or semi-analytic model which tracks stellar particles and has a coarser resolution than the FIRE simulations.\par
As an example application, we use our GMDN approach to post-process HII region line signals on top of well-resolved galaxies in the TNG50, TNG100, and TNG300 simulations at $z=6$. We present detailed comparisons between the models and observations as well as with other modeling efforts, considering
the galaxy-wide line emission signals, correlations between line luminosity and SFR, and galaxy line luminosity functions. 
We find that the multi-line GMDN $L(\mathrm{SFR})$ relationships differ from those in other work in the current literature. The differences partly trace to variations in the small-scale ISM properties across different galaxies, which are captured in our scheme, but neglected in most other current studies.
These differences highlight the necessity of a multi-scale simulation modeling framework for these lines, as important small-scale ISM property variations are otherwise neglected when predicting line emission statistics from large volume cosmological simulations. Although our GMDN model generally agrees with current observational measurements, it tends to under-predict the [OIII] 4364\AA\ line luminosity.
This line is highly sensitive to the HII region gas temperatures, and future versions of our model may benefit from improvements in our temperature models.\par
Although we find excellent agreement between simulated and observed line luminosity measurements, and line luminosity versus SFR relations, the simulated [OIII], H$\alpha$, and H$\beta$ line luminosity functions fall consistently lower than current measurements. We find that these discrepancies may owe to cosmic variance in the small fields spanned by current line luminosity function measurements.\par 
In summary, our machine learning based modeling framework is useful for exploring how galaxy formation model variations influence the observable ISM emission line statistics. 
By combining the galaxy population and stellar particle statistics sampled across large-volume cosmological simulations, with zoom-in galaxy simulations capable of resolving HII region properties, we avoid some of the simplifying assumptions made in previous work. While our current study focused entirely on simulated galaxies at $z=6$, and further studies are required to quantify the generality of our GMDN model, we expect this framework to be applicable to galaxies at $z\gtrsim6$. In fact, it may not be necessary to even re-train the GMDN provided the ISM properties evolve little across the relevant redshifts (e.g. $6 \lesssim z \lesssim 9$), as suggested by current observations and simulations including those related to the MZR, $L(\mathrm{SFR})$, and temperature versus metallicity relations (e.g. \citealt{2023MNRAS.525.5989Y,2024ApJ...967L..41M,2025arXiv250418006Y}).
We nevertheless aim to improve and extend our analysis in the near future. First, we hope to refine our treatment of the gas temperature in HII/OIII/OII regions. Second, we hope to extend the FIRE$+$\textsc{HIILines} calculations and GMDN modeling to lower redshifts to help interpret emission line observations at Cosmic Noon and in the more nearby universe. Third, it will be useful to extend \textsc{HIILines} to cover other HII region lines including OII recombination lines, [NII], [SII], and [SIII] lines, some of which will be observed by ALMA, the Roman Space Telescope, and multiple LIM surveys. Finally, note that we neglect contributions from Active Galactic Nuclei (AGN) in this work. In future studies, it would be interesting to include models for the line emission from gas exposed to ionizing radiation from an AGN within our framework. \par

\section{Acknowledgements}
S.Y. thanks Jialian Li for beneficial discussions about generative machine learning methods. S.Y. acknowledges support from the Director’s postdoctoral fellowship funded by LANL LDRD project No. 20240863PRD2. AL acknowledges support through NASA ATP grant 80NSSC20K0497. H.L. acknowledges the support by LANL/LDRD program.

\bibliography{MDN}{} 

\begin{thebibliography}{}
\expandafter\ifx\csname natexlab\endcsname\relax\def\natexlab#1{#1}\fi
\providecommand{\url}[1]{\href{#1}{#1}}
\providecommand{\dodoi}[1]{doi:~\href{http://doi.org/#1}{\nolinkurl{#1}}}
\providecommand{\doeprint}[1]{\href{http://ascl.net/#1}{\nolinkurl{http://ascl.net/#1}}}
\providecommand{\doarXiv}[1]{\href{https://arxiv.org/abs/#1}{\nolinkurl{https://arxiv.org/abs/#1}}}

\bibitem[{{Bernal} \& {Kovetz}(2022)}]{2022A&ARv..30....5B}
{Bernal}, J.~L., \& {Kovetz}, E.~D. 2022, \aapr, 30, 5, \dodoi{10.1007/s00159-022-00143-0}

\bibitem[{Bishop(1994)}]{Bishop1994MixtureDN}
Bishop, C.~M. 1994.
\newblock \url{https://api.semanticscholar.org/CorpusID:118227751}

\bibitem[{{Cardelli} {et~al.}(1989){Cardelli}, {Clayton}, \& {Mathis}}]{1989ApJ...345..245C}
{Cardelli}, J.~A., {Clayton}, G.~C., \& {Mathis}, J.~S. 1989, \apj, 345, 245, \dodoi{10.1086/167900}

\bibitem[{{Ceverino} {et~al.}(2021){Ceverino}, {Hirschmann}, {Klessen}, {Glover}, {Charlot}, \& {Feltre}}]{2021MNRAS.504.4472C}
{Ceverino}, D., {Hirschmann}, M., {Klessen}, R.~S., {et~al.} 2021, \mnras, 504, 4472, \dodoi{10.1093/mnras/stab1206}

\bibitem[{{Conroy} \& {Gunn}(2010)}]{2010ApJ...712..833C}
{Conroy}, C., \& {Gunn}, J.~E. 2010, \apj, 712, 833, \dodoi{10.1088/0004-637X/712/2/833}

\bibitem[{{Conroy} {et~al.}(2009){Conroy}, {Gunn}, \& {White}}]{2009ApJ...699..486C}
{Conroy}, C., {Gunn}, J.~E., \& {White}, M. 2009, \apj, 699, 486, \dodoi{10.1088/0004-637X/699/1/486}

\bibitem[{{Curti} {et~al.}(2022){Curti}, {D'Eugenio}, {Carniani}, {Maiolino}, {Sandles}, {Witstok}, {Baker}, {Bennett}, {Piotrowska}, {Tacchella}, {Charlot}, {Nakajima}, {Maheson}, {Mannucci}, {Arribas}, {Belfiore}, {Bonaventura}, {Bunker}, {Chevallard}, {Cresci}, {Curtis-Lake}, {Hayden-Pawson}, {Kumari}, {Laseter}, {Looser}, {Marconi}, {Maseda}, {Jones}, {Scholtz}, {Smit}, {Ubler}, \& {Wallace}}]{2022arXiv220712375C}
{Curti}, M., {D'Eugenio}, F., {Carniani}, S., {et~al.} 2022, arXiv e-prints, arXiv:2207.12375.
\newblock \doarXiv{2207.12375}

\bibitem[{{Curti} {et~al.}(2023){Curti}, {D'Eugenio}, {Carniani}, {Maiolino}, {Sandles}, {Witstok}, {Baker}, {Bennett}, {Piotrowska}, {Tacchella}, {Charlot}, {Nakajima}, {Maheson}, {Mannucci}, {Amiri}, {Arribas}, {Belfiore}, {Bonaventura}, {Bunker}, {Chevallard}, {Cresci}, {Curtis-Lake}, {Hayden-Pawson}, {Jones}, {Kumari}, {Laseter}, {Looser}, {Marconi}, {Maseda}, {Scholtz}, {Smit}, {{\"U}bler}, \& {Wallace}}]{2023MNRAS.518..425C}
---. 2023, \mnras, 518, 425, \dodoi{10.1093/mnras/stac2737}

\bibitem[{{De Barros} {et~al.}(2019){De Barros}, {Oesch}, {Labb{\'e}}, {Stefanon}, {Gonz{\'a}lez}, {Smit}, {Bouwens}, \& {Illingworth}}]{2019MNRAS.489.2355D}
{De Barros}, S., {Oesch}, P.~A., {Labb{\'e}}, I., {et~al.} 2019, \mnras, 489, 2355, \dodoi{10.1093/mnras/stz940}

\bibitem[{{Dor{\'e}} {et~al.}(2014){Dor{\'e}}, {Bock}, {Ashby}, {Capak}, {Cooray}, {de Putter}, {Eifler}, {Flagey}, {Gong}, {Habib}, {Heitmann}, {Hirata}, {Jeong}, {Katti}, {Korngut}, {Krause}, {Lee}, {Masters}, {Mauskopf}, {Melnick}, {Mennesson}, {Nguyen}, {{\"O}berg}, {Pullen}, {Raccanelli}, {Smith}, {Song}, {Tolls}, {Unwin}, {Venumadhav}, {Viero}, {Werner}, \& {Zemcov}}]{2014arXiv1412.4872D}
{Dor{\'e}}, O., {Bock}, J., {Ashby}, M., {et~al.} 2014, arXiv e-prints, arXiv:1412.4872.
\newblock \doarXiv{1412.4872}

\bibitem[{{Draine}(2011)}]{2011piim.book.....D}
{Draine}, B.~T. 2011, {Physics of the Interstellar and Intergalactic Medium}

\bibitem[{{Driver} \& {Robotham}(2010)}]{2010MNRAS.407.2131D}
{Driver}, S.~P., \& {Robotham}, A. S.~G. 2010, \mnras, 407, 2131, \dodoi{10.1111/j.1365-2966.2010.17028.x}

\bibitem[{{Ferland} {et~al.}(2017){Ferland}, {Chatzikos}, {Guzm{\'a}n}, {Lykins}, {van Hoof}, {Williams}, {Abel}, {Badnell}, {Keenan}, {Porter}, \& {Stancil}}]{2017RMxAA..53..385F}
{Ferland}, G.~J., {Chatzikos}, M., {Guzm{\'a}n}, F., {et~al.} 2017, \rmxaa, 53, 385.
\newblock \doarXiv{1705.10877}

\bibitem[{{Gong} {et~al.}(2017){Gong}, {Cooray}, {Silva}, {Zemcov}, {Feng}, {Santos}, {Dore}, \& {Chen}}]{2017ApJ...835..273G}
{Gong}, Y., {Cooray}, A., {Silva}, M.~B., {et~al.} 2017, \apj, 835, 273, \dodoi{10.3847/1538-4357/835/2/273}

\bibitem[{{Harikane} {et~al.}(2020){Harikane}, {Ouchi}, {Inoue}, {Matsuoka}, {Tamura}, {Bakx}, {Fujimoto}, {Moriwaki}, {Ono}, {Nagao}, {Tadaki}, {Kojima}, {Shibuya}, {Egami}, {Ferrara}, {Gallerani}, {Hashimoto}, {Kohno}, {Matsuda}, {Matsuo}, {Pallottini}, {Sugahara}, \& {Vallini}}]{2020ApJ...896...93H}
{Harikane}, Y., {Ouchi}, M., {Inoue}, A.~K., {et~al.} 2020, \apj, 896, 93, \dodoi{10.3847/1538-4357/ab94bd}

\bibitem[{{Hashimoto} {et~al.}(2018){Hashimoto}, {Laporte}, {Mawatari}, {Ellis}, {Inoue}, {Zackrisson}, {Roberts-Borsani}, {Zheng}, {Tamura}, {Bauer}, {Fletcher}, {Harikane}, {Hatsukade}, {Hayatsu}, {Matsuda}, {Matsuo}, {Okamoto}, {Ouchi}, {Pell{\'o}}, {Rydberg}, {Shimizu}, {Taniguchi}, {Umehata}, \& {Yoshida}}]{2018Natur.557..392H}
{Hashimoto}, T., {Laporte}, N., {Mawatari}, K., {et~al.} 2018, \nat, 557, 392, \dodoi{10.1038/s41586-018-0117-z}

\bibitem[{{Hashimoto} {et~al.}(2019){Hashimoto}, {Inoue}, {Mawatari}, {Tamura}, {Matsuo}, {Furusawa}, {Harikane}, {Shibuya}, {Knudsen}, {Kohno}, {Ono}, {Zackrisson}, {Okamoto}, {Kashikawa}, {Oesch}, {Ouchi}, {Ota}, {Shimizu}, {Taniguchi}, {Umehata}, \& {Watson}}]{2019PASJ...71...71H}
{Hashimoto}, T., {Inoue}, A.~K., {Mawatari}, K., {et~al.} 2019, \pasj, 71, 71, \dodoi{10.1093/pasj/psz049}

\bibitem[{{Heintz} {et~al.}(2022){Heintz}, {Brammer}, {Gim{\'e}nez-Arteaga}, {Strait}, {Lagos}, {Vijayan}, {Matthee}, {Watson}, {Mason}, {Hutter}, {Toft}, {Fynbo}, \& {Oesch}}]{2022arXiv221202890H}
{Heintz}, K.~E., {Brammer}, G.~B., {Gim{\'e}nez-Arteaga}, C., {et~al.} 2022, arXiv e-prints, arXiv:2212.02890, \dodoi{10.48550/arXiv.2212.02890}

\bibitem[{{Henry} {et~al.}(2021){Henry}, {Rafelski}, {Sunnquist}, {Pirzkal}, {Pacifici}, {Atek}, {Bagley}, {Baronchelli}, {Barro}, {Bunker}, {Colbert}, {Dai}, {Elmegreen}, {Elmegreen}, {Finkelstein}, {Kocevski}, {Koekemoer}, {Malkan}, {Martin}, {Mehta}, {Pahl}, {Papovich}, {Rutkowski}, {S{\'a}nchez Almeida}, {Scarlata}, {Snyder}, \& {Teplitz}}]{2021ApJ...919..143H}
{Henry}, A., {Rafelski}, M., {Sunnquist}, B., {et~al.} 2021, \apj, 919, 143, \dodoi{10.3847/1538-4357/ac1105}

\bibitem[{{Hirschmann} {et~al.}(2017){Hirschmann}, {Charlot}, {Feltre}, {Naab}, {Choi}, {Ostriker}, \& {Somerville}}]{2017MNRAS.472.2468H}
{Hirschmann}, M., {Charlot}, S., {Feltre}, A., {et~al.} 2017, \mnras, 472, 2468, \dodoi{10.1093/mnras/stx2180}

\bibitem[{{Hirschmann} {et~al.}(2023){Hirschmann}, {Charlot}, {Feltre}, {Curtis-Lake}, {Somerville}, {Chevallard}, {Choi}, {Nelson}, {Morisset}, {Plat}, \& {Vidal-Garcia}}]{2023MNRAS.526.3610H}
---. 2023, \mnras, 526, 3610, \dodoi{10.1093/mnras/stad2955}

\bibitem[{{Holweger}(2001)}]{2001AIPC..598...23H}
{Holweger}, H. 2001, in American Institute of Physics Conference Series, Vol. 598, Joint SOHO/ACE workshop ``Solar and Galactic Composition'', ed. R.~F. {Wimmer-Schweingruber}, 23--30, \dodoi{10.1063/1.1433974}

\bibitem[{{Inoue} {et~al.}(2016){Inoue}, {Tamura}, {Matsuo}, {Mawatari}, {Shimizu}, {Shibuya}, {Ota}, {Yoshida}, {Zackrisson}, {Kashikawa}, {Kohno}, {Umehata}, {Hatsukade}, {Iye}, {Matsuda}, {Okamoto}, \& {Yamaguchi}}]{2016Sci...352.1559I}
{Inoue}, A.~K., {Tamura}, Y., {Matsuo}, H., {et~al.} 2016, Science, 352, 1559, \dodoi{10.1126/science.aaf0714}

\bibitem[{{Kannan} {et~al.}(2022){Kannan}, {Smith}, {Garaldi}, {Shen}, {Vogelsberger}, {Pakmor}, {Springel}, \& {Hernquist}}]{2022MNRAS.514.3857K}
{Kannan}, R., {Smith}, A., {Garaldi}, E., {et~al.} 2022, \mnras, 514, 3857, \dodoi{10.1093/mnras/stac1557}

\bibitem[{{Katz} {et~al.}(2022){Katz}, {Rosdahl}, {Kimm}, {Garel}, {Blaizot}, {Haehnelt}, {Michel-Dansac}, {Martin-Alvarez}, {Devriendt}, {Slyz}, {Teyssier}, {Ocvirk}, {Laporte}, \& {Ellis}}]{2022MNRAS.510.5603K}
{Katz}, H., {Rosdahl}, J., {Kimm}, T., {et~al.} 2022, \mnras, 510, 5603, \dodoi{10.1093/mnras/stac028}

\bibitem[{{Kingma} \& {Ba}(2014)}]{2014arXiv1412.6980K}
{Kingma}, D.~P., \& {Ba}, J. 2014, arXiv e-prints, arXiv:1412.6980, \dodoi{10.48550/arXiv.1412.6980}

\bibitem[{{Laporte} {et~al.}(2017){Laporte}, {Ellis}, {Boone}, {Bauer}, {Qu{\'e}nard}, {Roberts-Borsani}, {Pell{\'o}}, {P{\'e}rez-Fournon}, \& {Streblyanska}}]{2017ApJ...837L..21L}
{Laporte}, N., {Ellis}, R.~S., {Boone}, F., {et~al.} 2017, \apjl, 837, L21, \dodoi{10.3847/2041-8213/aa62aa}

\bibitem[{{Laporte} {et~al.}(2019){Laporte}, {Katz}, {Ellis}, {Lagache}, {Bauer}, {Boone}, {Inoue}, {Hashimoto}, {Matsuo}, {Mawatari}, \& {Tamura}}]{2019MNRAS.487L..81L}
{Laporte}, N., {Katz}, H., {Ellis}, R.~S., {et~al.} 2019, \mnras, 487, L81, \dodoi{10.1093/mnrasl/slz094}

\bibitem[{{Laseter} {et~al.}(2023){Laseter}, {Maseda}, {Curti}, {Maiolino}, {D'Eugenio}, {Cameron}, {Looser}, {Arribas}, {Baker}, {Bhatawdekar}, {Boyett}, {Bunker}, {Carniani}, {Charlot}, {Chevallard}, {Curtis-lake}, {Egami}, {Eisenstein}, {Hainline}, {Hausen}, {Ji}, {Kumari}, {Perna}, {Rawle}, {Rix}, {Robertson}, {Rodr{\'\i}guez Del Pino}, {Sandles}, {Scholtz}, {Smit}, {Tacchella}, {{\"U}bler}, {Williams}, {Willott}, \& {Witstok}}]{2023arXiv230603120L}
{Laseter}, I.~H., {Maseda}, M.~V., {Curti}, M., {et~al.} 2023, arXiv e-prints, arXiv:2306.03120, \dodoi{10.48550/arXiv.2306.03120}

\bibitem[{{Lee} {et~al.}(2006){Lee}, {Skillman}, {Cannon}, {Jackson}, {Gehrz}, {Polomski}, \& {Woodward}}]{2006ApJ...647..970L}
{Lee}, H., {Skillman}, E.~D., {Cannon}, J.~M., {et~al.} 2006, \apj, 647, 970, \dodoi{10.1086/505573}

\bibitem[{{Ma} {et~al.}(2018){Ma}, {Hopkins}, {Garrison-Kimmel}, {Faucher-Gigu{\`e}re}, {Quataert}, {Boylan-Kolchin}, {Hayward}, {Feldmann}, \& {Kere{\v{s}}}}]{2018MNRAS.478.1694M}
{Ma}, X., {Hopkins}, P.~F., {Garrison-Kimmel}, S., {et~al.} 2018, \mnras, 478, 1694, \dodoi{10.1093/mnras/sty1024}

\bibitem[{{Ma} {et~al.}(2019){Ma}, {Hayward}, {Casey}, {Hopkins}, {Quataert}, {Liang}, {Faucher-Gigu{\`e}re}, {Feldmann}, \& {Kere{\v{s}}}}]{2019MNRAS.487.1844M}
{Ma}, X., {Hayward}, C.~C., {Casey}, C.~M., {et~al.} 2019, \mnras, 487, 1844, \dodoi{10.1093/mnras/stz1324}

\bibitem[{{Ma} {et~al.}(2020){Ma}, {Grudi{\'c}}, {Quataert}, {Hopkins}, {Faucher-Gigu{\`e}re}, {Boylan-Kolchin}, {Wetzel}, {Kim}, {Murray}, \& {Kere{\v{s}}}}]{2020MNRAS.493.4315M}
{Ma}, X., {Grudi{\'c}}, M.~Y., {Quataert}, E., {et~al.} 2020, \mnras, 493, 4315, \dodoi{10.1093/mnras/staa527}

\bibitem[{{Maiolino} {et~al.}(2008){Maiolino}, {Nagao}, {Grazian}, {Cocchia}, {Marconi}, {Mannucci}, {Cimatti}, {Pipino}, {Ballero}, {Calura}, {Chiappini}, {Fontana}, {Granato}, {Matteucci}, {Pastorini}, {Pentericci}, {Risaliti}, {Salvati}, \& {Silva}}]{2008A&A...488..463M}
{Maiolino}, R., {Nagao}, T., {Grazian}, A., {et~al.} 2008, \aap, 488, 463, \dodoi{10.1051/0004-6361:200809678}

\bibitem[{{Mannucci} {et~al.}(2009){Mannucci}, {Cresci}, {Maiolino}, {Marconi}, {Pastorini}, {Pozzetti}, {Gnerucci}, {Risaliti}, {Schneider}, {Lehnert}, \& {Salvati}}]{2009MNRAS.398.1915M}
{Mannucci}, F., {Cresci}, G., {Maiolino}, R., {et~al.} 2009, \mnras, 398, 1915, \dodoi{10.1111/j.1365-2966.2009.15185.x}

\bibitem[{{Marinacci} {et~al.}(2018){Marinacci}, {Vogelsberger}, {Pakmor}, {Torrey}, {Springel}, {Hernquist}, {Nelson}, {Weinberger}, {Pillepich}, {Naiman}, \& {Genel}}]{2018MNRAS.480.5113M}
{Marinacci}, F., {Vogelsberger}, M., {Pakmor}, R., {et~al.} 2018, \mnras, 480, 5113, \dodoi{10.1093/mnras/sty2206}

\bibitem[{{Marszewski} {et~al.}(2024){Marszewski}, {Sun}, {Faucher-Gigu{\`e}re}, {Hayward}, \& {Feldmann}}]{2024ApJ...967L..41M}
{Marszewski}, A., {Sun}, G., {Faucher-Gigu{\`e}re}, C.-A., {Hayward}, C.~C., \& {Feldmann}, R. 2024, \apjl, 967, L41, \dodoi{10.3847/2041-8213/ad4cee}

\bibitem[{{Matthee} {et~al.}(2023){Matthee}, {Mackenzie}, {Simcoe}, {Kashino}, {Lilly}, {Bordoloi}, \& {Eilers}}]{2023ApJ...950...67M}
{Matthee}, J., {Mackenzie}, R., {Simcoe}, R.~A., {et~al.} 2023, \apj, 950, 67, \dodoi{10.3847/1538-4357/acc846}

\bibitem[{{Moriwaki} {et~al.}(2018){Moriwaki}, {Yoshida}, {Shimizu}, {Harikane}, {Matsuda}, {Matsuo}, {Hashimoto}, {Inoue}, {Tamura}, \& {Nagao}}]{2018MNRAS.481L..84M}
{Moriwaki}, K., {Yoshida}, N., {Shimizu}, I., {et~al.} 2018, \mnras, 481, L84, \dodoi{10.1093/mnrasl/sly167}

\bibitem[{{Naiman} {et~al.}(2018){Naiman}, {Pillepich}, {Springel}, {Ramirez-Ruiz}, {Torrey}, {Vogelsberger}, {Pakmor}, {Nelson}, {Marinacci}, {Hernquist}, {Weinberger}, \& {Genel}}]{2018MNRAS.477.1206N}
{Naiman}, J.~P., {Pillepich}, A., {Springel}, V., {et~al.} 2018, \mnras, 477, 1206, \dodoi{10.1093/mnras/sty618}

\bibitem[{{Nakajima} {et~al.}(2023){Nakajima}, {Ouchi}, {Isobe}, {Harikane}, {Zhang}, {Ono}, {Umeda}, \& {Oguri}}]{2023ApJS..269...33N}
{Nakajima}, K., {Ouchi}, M., {Isobe}, Y., {et~al.} 2023, \apjs, 269, 33, \dodoi{10.3847/1538-4365/acd556}

\bibitem[{{Nakazato} {et~al.}(2023){Nakazato}, {Yoshida}, \& {Ceverino}}]{2023ApJ...953..140N}
{Nakazato}, Y., {Yoshida}, N., \& {Ceverino}, D. 2023, \apj, 953, 140, \dodoi{10.3847/1538-4357/ace25a}

\bibitem[{{Nelson} {et~al.}(2018){Nelson}, {Pillepich}, {Springel}, {Weinberger}, {Hernquist}, {Pakmor}, {Genel}, {Torrey}, {Vogelsberger}, {Kauffmann}, {Marinacci}, \& {Naiman}}]{2018MNRAS.475..624N}
{Nelson}, D., {Pillepich}, A., {Springel}, V., {et~al.} 2018, \mnras, 475, 624, \dodoi{10.1093/mnras/stx3040}

\bibitem[{{Nelson} {et~al.}(2019{\natexlab{a}}){Nelson}, {Pillepich}, {Springel}, {Pakmor}, {Weinberger}, {Genel}, {Torrey}, {Vogelsberger}, {Marinacci}, \& {Hernquist}}]{2019MNRAS.490.3234N}
---. 2019{\natexlab{a}}, \mnras, 490, 3234, \dodoi{10.1093/mnras/stz2306}

\bibitem[{{Nelson} {et~al.}(2019{\natexlab{b}}){Nelson}, {Springel}, {Pillepich}, {Rodriguez-Gomez}, {Torrey}, {Genel}, {Vogelsberger}, {Pakmor}, {Marinacci}, {Weinberger}, {Kelley}, {Lovell}, {Diemer}, \& {Hernquist}}]{2019ComAC...6....2N}
{Nelson}, D., {Springel}, V., {Pillepich}, A., {et~al.} 2019{\natexlab{b}}, Computational Astrophysics and Cosmology, 6, 2, \dodoi{10.1186/s40668-019-0028-x}

\bibitem[{{Osterbrock} \& {Ferland}(2006)}]{2006agna.book.....O}
{Osterbrock}, D.~E., \& {Ferland}, G.~J. 2006, {Astrophysics of gaseous nebulae and active galactic nuclei}

\bibitem[{{Padmanabhan}(2018)}]{2018MNRAS.475.1477P}
{Padmanabhan}, H. 2018, \mnras, 475, 1477, \dodoi{10.1093/mnras/stx3250}

\bibitem[{{Pillepich} {et~al.}(2018{\natexlab{a}}){Pillepich}, {Springel}, {Nelson}, {Genel}, {Naiman}, {Pakmor}, {Hernquist}, {Torrey}, {Vogelsberger}, {Weinberger}, \& {Marinacci}}]{2018MNRAS.473.4077P}
{Pillepich}, A., {Springel}, V., {Nelson}, D., {et~al.} 2018{\natexlab{a}}, \mnras, 473, 4077, \dodoi{10.1093/mnras/stx2656}

\bibitem[{{Pillepich} {et~al.}(2018{\natexlab{b}}){Pillepich}, {Nelson}, {Hernquist}, {Springel}, {Pakmor}, {Torrey}, {Weinberger}, {Genel}, {Naiman}, {Marinacci}, \& {Vogelsberger}}]{2018MNRAS.475..648P}
{Pillepich}, A., {Nelson}, D., {Hernquist}, L., {et~al.} 2018{\natexlab{b}}, \mnras, 475, 648, \dodoi{10.1093/mnras/stx3112}

\bibitem[{{Pillepich} {et~al.}(2019){Pillepich}, {Nelson}, {Springel}, {Pakmor}, {Torrey}, {Weinberger}, {Vogelsberger}, {Marinacci}, {Genel}, {van der Wel}, \& {Hernquist}}]{2019MNRAS.490.3196P}
{Pillepich}, A., {Nelson}, D., {Springel}, V., {et~al.} 2019, \mnras, 490, 3196, \dodoi{10.1093/mnras/stz2338}

\bibitem[{{Popping} {et~al.}(2014){Popping}, {Somerville}, \& {Trager}}]{2014MNRAS.442.2398P}
{Popping}, G., {Somerville}, R.~S., \& {Trager}, S.~C. 2014, \mnras, 442, 2398, \dodoi{10.1093/mnras/stu991}

\bibitem[{{Porter} {et~al.}(2014){Porter}, {Somerville}, {Primack}, \& {Johansson}}]{2014MNRAS.444..942P}
{Porter}, L.~A., {Somerville}, R.~S., {Primack}, J.~R., \& {Johansson}, P.~H. 2014, \mnras, 444, 942, \dodoi{10.1093/mnras/stu1434}

\bibitem[{{Sanders} {et~al.}(2023){Sanders}, {Shapley}, {Topping}, {Reddy}, \& {Brammer}}]{2023arXiv230308149S}
{Sanders}, R.~L., {Shapley}, A.~E., {Topping}, M.~W., {Reddy}, N.~A., \& {Brammer}, G.~B. 2023, arXiv e-prints, arXiv:2303.08149, \dodoi{10.48550/arXiv.2303.08149}

\bibitem[{{Schaerer}(2003)}]{2003A&A...397..527S}
{Schaerer}, D. 2003, \aap, 397, 527, \dodoi{10.1051/0004-6361:20021525}

\bibitem[{{Somerville} {et~al.}(2012){Somerville}, {Gilmore}, {Primack}, \& {Dom{\'\i}nguez}}]{2012MNRAS.423.1992S}
{Somerville}, R.~S., {Gilmore}, R.~C., {Primack}, J.~R., \& {Dom{\'\i}nguez}, A. 2012, \mnras, 423, 1992, \dodoi{10.1111/j.1365-2966.2012.20490.x}

\bibitem[{{Somerville} {et~al.}(2008){Somerville}, {Hopkins}, {Cox}, {Robertson}, \& {Hernquist}}]{2008MNRAS.391..481S}
{Somerville}, R.~S., {Hopkins}, P.~F., {Cox}, T.~J., {Robertson}, B.~E., \& {Hernquist}, L. 2008, \mnras, 391, 481, \dodoi{10.1111/j.1365-2966.2008.13805.x}

\bibitem[{{Somerville} {et~al.}(2015){Somerville}, {Popping}, \& {Trager}}]{2015MNRAS.453.4337S}
{Somerville}, R.~S., {Popping}, G., \& {Trager}, S.~C. 2015, \mnras, 453, 4337, \dodoi{10.1093/mnras/stv1877}

\bibitem[{{Somerville} \& {Primack}(1999)}]{1999MNRAS.310.1087S}
{Somerville}, R.~S., \& {Primack}, J.~R. 1999, \mnras, 310, 1087, \dodoi{10.1046/j.1365-8711.1999.03032.x}

\bibitem[{{Springel} \& {Hernquist}(2003)}]{Springel2003}
{Springel}, V., \& {Hernquist}, L. 2003, \mnras, 339, 289, \dodoi{10.1046/j.1365-8711.2003.06206.x}

\bibitem[{{Springel} {et~al.}(2018){Springel}, {Pakmor}, {Pillepich}, {Weinberger}, {Nelson}, {Hernquist}, {Vogelsberger}, {Genel}, {Torrey}, {Marinacci}, \& {Naiman}}]{2018MNRAS.475..676S}
{Springel}, V., {Pakmor}, R., {Pillepich}, A., {et~al.} 2018, \mnras, 475, 676, \dodoi{10.1093/mnras/stx3304}

\bibitem[{{Sun} {et~al.}(2023{\natexlab{a}}){Sun}, {Egami}, {Pirzkal}, {Rieke}, {Baum}, {Boyer}, {Boyett}, {Bunker}, {Cameron}, {Curti}, {Eisenstein}, {Gennaro}, {Greene}, {Jaffe}, {Kelly}, {Koekemoer}, {Kumari}, {Maiolino}, {Maseda}, {Perna}, {Rest}, {Robertson}, {Schlawin}, {Smit}, {Stansberry}, {Sunnquist}, {Tacchella}, {Williams}, \& {Willmer}}]{2023ApJ...953...53S}
{Sun}, F., {Egami}, E., {Pirzkal}, N., {et~al.} 2023{\natexlab{a}}, \apj, 953, 53, \dodoi{10.3847/1538-4357/acd53c}

\bibitem[{{Sun} {et~al.}(2023{\natexlab{b}}){Sun}, {Faucher-Gigu{\`e}re}, {Hayward}, {Shen}, {Wetzel}, \& {Cochrane}}]{2023ApJ...955L..35S}
{Sun}, G., {Faucher-Gigu{\`e}re}, C.-A., {Hayward}, C.~C., {et~al.} 2023{\natexlab{b}}, \apjl, 955, L35, \dodoi{10.3847/2041-8213/acf85a}

\bibitem[{{Sun} {et~al.}(2021){Sun}, {Chang}, {Uzgil}, {Bock}, {Bradford}, {Butler}, {Caze-Cortes}, {Cheng}, {Cooray}, {Crites}, {Hailey-Dunsheath}, {Emerson}, {Frez}, {Hoscheit}, {Hunacek}, {Keenan}, {Li}, {Madonia}, {Marrone}, {Moncelsi}, {Shiu}, {Trumper}, {Turner}, {Weber}, {Wei}, \& {Zemcov}}]{2021ApJ...915...33S}
{Sun}, G., {Chang}, T.~C., {Uzgil}, B.~D., {et~al.} 2021, \apj, 915, 33, \dodoi{10.3847/1538-4357/abfe62}

\bibitem[{{Tremonti} {et~al.}(2004){Tremonti}, {Heckman}, {Kauffmann}, {Brinchmann}, {Charlot}, {White}, {Seibert}, {Peng}, {Schlegel}, {Uomoto}, {Fukugita}, \& {Brinkmann}}]{2004ApJ...613..898T}
{Tremonti}, C.~A., {Heckman}, T.~M., {Kauffmann}, G., {et~al.} 2004, \apj, 613, 898, \dodoi{10.1086/423264}

\bibitem[{{Vogelsberger} {et~al.}(2020){Vogelsberger}, {Nelson}, {Pillepich}, {Shen}, {Marinacci}, {Springel}, {Pakmor}, {Tacchella}, {Weinberger}, {Torrey}, \& {Hernquist}}]{2020MNRAS.492.5167V}
{Vogelsberger}, M., {Nelson}, D., {Pillepich}, A., {et~al.} 2020, \mnras, 492, 5167, \dodoi{10.1093/mnras/staa137}

\bibitem[{{Witstok} {et~al.}(2022){Witstok}, {Smit}, {Maiolino}, {Kumari}, {Aravena}, {Boogaard}, {Bouwens}, {Carniani}, {Hodge}, {Jones}, {Stefanon}, {van der Werf}, \& {Schouws}}]{2022MNRAS.515.1751W}
{Witstok}, J., {Smit}, R., {Maiolino}, R., {et~al.} 2022, \mnras, 515, 1751, \dodoi{10.1093/mnras/stac1905}

\bibitem[{{Yang} \& {Lidz}(2020)}]{2020MNRAS.499.3417Y}
{Yang}, S., \& {Lidz}, A. 2020, \mnras, 499, 3417, \dodoi{10.1093/mnras/staa3000}

\bibitem[{{Yang} {et~al.}(2023{\natexlab{a}}){Yang}, {Lidz}, {Benson}, {Singh Chauhan}, {Smith}, \& {Li}}]{2023arXiv231209213Y}
{Yang}, S., {Lidz}, A., {Benson}, A., {et~al.} 2023{\natexlab{a}}, arXiv e-prints, arXiv:2312.09213, \dodoi{10.48550/arXiv.2312.09213}

\bibitem[{{Yang} {et~al.}(2025){Yang}, {Lidz}, {Li}, {Popping}, {Zavala}, \& {Sun}}]{2025arXiv250418006Y}
{Yang}, S., {Lidz}, A., {Li}, H., {et~al.} 2025, arXiv e-prints, arXiv:2504.18006.
\newblock \doarXiv{2504.18006}

\bibitem[{{Yang} {et~al.}(2023{\natexlab{b}}){Yang}, {Lidz}, {Smith}, {Benson}, \& {Li}}]{2023MNRAS.525.5989Y}
{Yang}, S., {Lidz}, A., {Smith}, A., {Benson}, A., \& {Li}, H. 2023{\natexlab{b}}, \mnras, 525, 5989, \dodoi{10.1093/mnras/stad2571}

\bibitem[{{Yang} {et~al.}(2022){Yang}, {Popping}, {Somerville}, {Pullen}, {Breysse}, \& {Maniyar}}]{2022ApJ...929..140Y}
{Yang}, S., {Popping}, G., {Somerville}, R.~S., {et~al.} 2022, \apj, 929, 140, \dodoi{10.3847/1538-4357/ac5d57}

\bibitem[{{Yates} {et~al.}(2020){Yates}, {Schady}, {Chen}, {Schweyer}, \& {Wiseman}}]{2020A&A...634A.107Y}
{Yates}, R.~M., {Schady}, P., {Chen}, T.~W., {Schweyer}, T., \& {Wiseman}, P. 2020, \aap, 634, A107, \dodoi{10.1051/0004-6361/201936506}

\bibitem[{{Zahid} {et~al.}(2013){Zahid}, {Geller}, {Kewley}, {Hwang}, {Fabricant}, \& {Kurtz}}]{2013ApJ...771L..19Z}
{Zahid}, H.~J., {Geller}, M.~J., {Kewley}, L.~J., {et~al.} 2013, \apjl, 771, L19, \dodoi{10.1088/2041-8205/771/2/L19}

\end{thebibliography}
\bibliographystyle{aasjournal}

\end{document}